\documentclass[preprint,prd,aps,floats,nofootinbib,floatfix]
{revtex4-2}

\renewcommand\i{\iota}

\newcommand{\diracslash}[1]{#1\llap{/\kern2pt}}

\newcommand{\be}{\begin{equation}}
\newcommand{\ee}{\end{equation}}
\newcommand{\bea}{\begin{eqnarray}}
\newcommand{\eea}{\end{eqnarray}}
\newcommand{\ba}[1]{\begin{array}{#1}}
\newcommand{\ea}{\end{array}}

\newcommand{\bt}{\begin{tabular}}
\newcommand{\et}{\end{tabular}}

\newcommand{\beas}{\begin{eqnarray*}}
\newcommand{\eeas}{\end{eqnarray*}}

\DeclareSymbolFont{rsfs}{U}{rsfs}{m}{n}
\DeclareSymbolFontAlphabet{\mathrsfs}{rsfs}

\usepackage{graphicx}

\usepackage{ulem}
\usepackage{multirow}
\usepackage{graphicx,epstopdf}
\usepackage{amsmath}
\usepackage{graphicx}
\usepackage{subcaption}
\usepackage{comment}
\usepackage[utf8]{inputenc}

\usepackage{csquotes}
\usepackage[english]{babel}
\usepackage{hyperref}
\usepackage{cleveref}
\hypersetup{
    colorlinks=true,
    linkcolor=blue,
    filecolor=magenta,      
    citecolor=blue, 
    urlcolor=blue,
}
 
\begin{document}


\title{Impact of finite volume on kaon, antikaon, and $\phi$ meson masses and decay width in asymmetric strange hadronic matter} 
 \author{Zeeshan Ahmad}
\email{zeeal143@hotmail.com }
\affiliation{Department of Physics, Dr. B R Ambedkar National Institute of Technology Jalandhar, 
 Jalandhar -- 144008, Punjab, India}

 \author{Nisha Chahal}
\email{nishachahal137@gmail.com}
\affiliation{Department of Physics, Dr. B R Ambedkar National Institute of Technology Jalandhar, 
 Jalandhar -- 144008, Punjab, India}

\author{Arvind Kumar}
\email{kumara@nitj.ac.in}
\affiliation{Department of Physics, Dr. B R Ambedkar National Institute of Technology Jalandhar, 
 Jalandhar -- 144008, Punjab, India}

\author{Suneel Dutt}
\email{dutts@nitj.ac.in}
\affiliation{Department of Physics, Dr. B R Ambedkar National Institute of Technology Jalandhar, 
 Jalandhar -- 144008, Punjab, India}

\def\be{\begin{equation}}
\def\ee{\end{equation}}
\def\bearr{\begin{eqnarray}}
\def\eearr{\end{eqnarray}}
\def\zbf#1{{\bf {#1}}}
\def\bfm#1{\mbox{\boldmath $#1$}}
\def\hf{\frac{1}{2}}
\def\kp{\zbf k+\frac{\zbf q}{2}}
\def\km{-\zbf k+\frac{\zbf q}{2}}
\def\hwo{\hat\omega_1}
\def\hwt{\hat\omega_2}

\begin{abstract}

In the present work, we investigate the impact of finite volume on the in-medium properties of kaons ($K^+$, $K^0$) and antikaons ($K^-$, $\bar{K^0}$), and $\phi$ mesons in the isospin asymmetric strange hadronic medium at finite density and temperature. 
We use the chiral SU(3) hadronic mean-field model, which accounts for the interactions between baryons through the exchange of scalar ($\sigma, \zeta, \delta $) and vector ($\omega$, $\rho$, $\phi$) fields. To investigate the effects of finite volume, we apply the multiple reflection expansion (MRE) technique for calculations of the density of states. The non-strange scalar field $\sigma$ shows significant variation in an asymmetric medium, while the strange scalar field $\zeta$ shows  good dependency in the strange medium. 
We use the medium-modified masses of kaons and antikaons calculated using the chiral SU(3) model to obtain the masses and decay width of $\phi$ mesons in finite volume hadronic medium.
To obtain the masses and decay widths of $\phi$ mesons, an effective Lagrangian approach 
  with  $\phi$$K$$\bar{K}$ interactions at one-loop level is used in the present work.
  We obtain the effective masses and decay widths in the finite volume matter, for the spherical geometry of the medium with Neumann and Dirichlet boundary conditions as well as for the cubic geometry. The finite volume effects are  found to be appreciable at high baryon densities. 

\end{abstract}

\maketitle

 \newpage

\section{Introduction}
\label{intro}
The study of the in-medium properties of hadrons at finite temperatures and densities continues to receive significant attention due to their significant role in understanding the strong interaction physics of the quantum chromodynamics (QCD) phase diagram  \cite{Gross:1980br,Mannarelli:2005pz,Leupold:2009kz,Haidenbauer:2018gvg,Petschauer:2015nea,Holzenkamp:1989tq,Tolos:2020aln,Lee:1987mj,Kim:2021xyp,Montana:2023sft,Kumar:2024owe}. At low temperatures and baryon chemical potential, colorless hadrons appear as essential constituents in the QCD   \cite{Ayala:2020rmb}. However, as the temperature and/or baryon density
of the medium is increased, a phase transition occurs, leading to the deconfinement of hadrons into quarks and gluons, and quark-gluon plasma (QGP) phase may form \cite{Aoki:2006br,Borsanyi:2010bp,Borsanyi:2015waa,Borsanyi:2013cga,Ding:2015ona,friman:2011cbm,Hatsuda:1994pi,Letessier:2002ony}. Lattice QCD uses a grid-based simulation technique to explore the phase diagram of QCD at low baryonic densities, shedding light on the behavior of quarks and gluons in strongly interacting medium \cite{Wilson:1974sk}.
  At the zero value of baryon chemical potential, lattice calculations  estimate the temperature for the chiral crossover from hadronic to QGP phase at around $T_c \approx 156$ MeV \cite{HotQCD:2018pds,
	Fischer:2018sdj,Aoki:2006we,Aoki:2006br}. Though the lattice simulations have been successful at the chemical potentials values close to vicinity of zero, but due to the sign problem, these techniques are not applied to study the phase transition at high baryon chemical potential values \cite{karsch:2002lattice, Borsanyi:2012cr}.


In the experimental facilities of  heavy-ion collisions (HICs) such as Large Hadron Collider (LHC) at CERN and Relativistic Heavy-Ion Collider (RHIC) at BNL, USA, hot and dense matter at high temperatures and low baryon chemical potential
has been produced \cite{Braun-Munzinger:2015hba},
 whereas the future Facility for Antiproton and Ion Research (FAIR) project at GSI-Germany, Nuclotron-based Ion Collider fAcility (NICA) at Dubna, and the Japan Proton Accelerator Research Complex (J-PARC) aim to examine the QCD phase diagram 
at moderate temperatures and high baryon densities \cite{Toneev:2007yu}. In the high-energy HICs, when two
isospin  asymmetric nuclei collide, the QGP fireball, which may be produced  initially, further transforms into a hadronic phase with the decrease in temperature.
The size of the fireball produced in these experiments is not of 
infinite extent, but has a finite size which depends
upon several factors such as the collisions centrality, size of colliding nuclei, and center of mass energy \cite{Graef:2012sh,Bzdak:2013zma,Hirono:2014dda}.

Numerous studies have been conducted to estimate the critical endpoint and for understanding the thermodynamics of strongly interacting matter, particularly focusing on the impact of  finite volume of
QCD matter
\cite{Fisher:1972zza, Bhattacharyya:2014uxa, Fraga:2011hi, Abreu:2006pt}. Studies have shown that the location of the critical endpoint and the behavior different thermodynamic quantities is notably influenced by consideration of the finite volume of the medium. According to Polyakov linear sigma model (PLSM) \cite{Bochkarev:1995gi,Mao:2009aq}, with decreasing system volume, the quark-hadron phase boundary is shifted towards higher values of $\mu$ and $T$ \cite{Magdy:2015eda}, while the critical endpoint shifts towards higher $\mu$ and lower $T$ \cite{Fraga:2011hi,Palhares:2009tf}. In theoretical computations, the volume of strongly interacting matter can be conceptualized with various sizes and shapes. Extensive studies have focused on examining the consequences of a finite volume on the QCD phase diagram using a cubic box \cite{Abreu:2019czp, Wang:2018qyq, Magdy:2019frj, Liu:2020elq}. However, some studies have considered the more realistic spherical shape \cite{Zhang:2019gva, Xu:2020loz}.
The multiple reflection expansion (MRE) method, which accounts for surface and curvature effects, refines the computation of the density of states by including essential corrections for small systems \cite{Zhao:2018nqa, Kiriyama:2005eh}. 
The effects of finite size have been explored through different approaches, such as chiral perturbation theory \cite{Hansen:1990un,Damgaard:2008zs}, the quark-meson model \cite{Braun:2004yk,Braun:2005fj,Colangelo:2005gd,Colangelo:2010ba}, the Dyson-Schwinger equations employing anti-periodic boundary conditions \cite{Shi:2018tsq}, the Polyakov loop extended Nambu–Jona-Lasinio (NJL) model \cite{Bhattacharyya:2014uxa,Pan:2016ecs}, and various non-perturbative renormalization group techniques \cite{Tripolt:2013zfa}.
The finite size effects on the properties of strongly interacting matter have been explored using NJL \cite{Abreu:2019czp,Abreu:2006pt} and relativistic mean field model \cite{Abreu:2017lxf,Abreu:2019tnf},  
applying anti-periodic boundary conditions for the fermions in the imaginary time coordinate, imposed via  Kubo-Martin Schwinger conditions. In the spatial coordinate directions, no restrictions on the boundary conditions were imposed, and choice of these led to significant observations on the masses of mesons. 
A comprehensive review of finite size effects is available in Ref. \cite{Klein:2017shl}.
The above-listed studies using different non-perturbative techniques have mostly considered the effects of finite size on the quark matter phase.
However, the impact of finite size effects on the physical
properties, which are calculated in hot and dense 
hadronic fireball has not been studied much.
In Ref. \cite{Samanta:2017ohm}, the transport coefficients of hadronic matter within the framework of the Hadron Resonance Gas (HRG) model have been studied employing lower momentum cut-off.

The  in-medium masses and decay width of different mesons at finite  baryon density and temperatures have been studied
using different approaches \cite{Hatsuda:1991ez,Hatsuda:1994pi,
Metag:2017yuh,Kim:2019ybi,
Ramos:1999ku,
Tsushima:1997df,
Mishra:2004te,Mishra:2003se,Kumar:2010gb,Kumar:2011ff,Mishra:2006wy,Zschiesche:2003qq,Hosaka:2016ypm,Bozkir:2022lyk,Yeo:2024iqz,Ruivo:1999pr}. 
The modifications of optical potentials play a significant role in understanding various experimental observables \cite{Feng:2015vra,Xu:2013aza,Paryev:2021duz}.
The investigations of the in-medium characteristics of kaons and antikaons were initiated by Kaplan and Nelson \cite{Kaplan:1986yq,Nelson:1987dg}.
The importance of $K$ and $\bar{K}$ isospin doublets in heavy-ion collisions has led to extensive theoretical \cite{Mishra:2008dj,Lutz:1997wt,Li:1997zb,Ko:2000cd,Cassing:1996xx,Bratkovskaya:1997pj,Cassing:1999es,Lutz:2001yb,Ramos:1999ku} and experimental \cite{Forster:2002sc,KaoS:2000eil,KaoS:1999gal} research. The low-energy antikaon–proton scattering data was successfully explained in terms of s-wave coupled SU(3) channels in \cite{Kaiser:1995eg,Oset:1997it}, one of the first interesting attempts in this area. 
 Moreover, the characteristics of $K$ and $\Bar{K}$ mesons are investigated in isospin asymmetric nuclear   \cite{Mishra:2008kg} and strange matter  \cite{Mishra:2008dj} using the chiral SU(3) model \cite{Papazoglou:1998vr}.
In the various non-perturbative  approaches described above to calculate the in-medium masses or decay width of different mesons, the dense hadronic medium is assumed to be of infinite extent.
It will be of interest and also important to examine the impact of finite volume of the hadronic phase on the optical potentials of different mesons. 
In the present work, we  investigate the in-medium masses of pseudoscalar kaons, $K\left(K^+, K^0\right)$ and antikaons $\bar{K} \left(K^{-}, \bar{K^0}\right)$ in a finite volume isospin asymmetric hot and dense medium composed of nucleons and hyperons.  Further, using these medium modified masses of $K$ and $\bar{K}$ mesons, we shall also study the in-medium masses and decay width of $\phi$ mesons in finite volume matter.

The motivation to investigate the in-medium properties of $\phi$ mesons is as follows. Light vector mesons ($\omega$, $\rho$, and $\phi$) play a crucial role in dilepton formation in HICs, prompting interest in both theoretical and experimental investigations \cite{Mishra:2014rha,Leupold:2009kz,Hayano:2008vn,Kumar:2018pqs,Cobos-Martinez:2017vtr, Kim:2022eku, Vujanovic:2008zf,Kumar:2020vys}. Because dileptons interact weakly with baryons and mesons, these are regarded as a promising observable to study the properties of hadronic medium \cite{Gale:1987ki,Xia:1988ym,Ko:1989vi,Korpa:1990ww,Gale:1997pe}.  
In the case of  $\phi$ mesons, despite being predominantly composed of strange quarks, they interact strongly with nuclei containing up and down quarks as well as with strange baryons and mesons \cite{Tsushima:1997df,Cabrera:2002hc, Cabrera:2016rnc, Cobos-Martinez:2017vtr}. 
Since almost 85\% of the $\phi$ meson decays to kaons hence, in-medium properties of kaons and antikaons play a crucial role in the determination of the properties of $\phi$ mesons.
For $\phi$ mesons, at normal nuclear matter density, the KEK-E325 collaboration \cite{KEK-PS-E325:2005wbm} observed an increase in the decay width by 3.6 times with a decrease in mass by 3.4\%. SPring8 facility  \cite{Ishikawa:2004id}, revealed a substantial in-medium $\phi$N cross-section in comparison to KEK-E325 collaboration, leading to a decay width of 35 MeV. These results are consistent with other experiments \cite{CLAS:2009kjz, CLAS:2007xhu}.
In an effort to shed light on the discrepancies, the CLAS collaboration at JLab conducted new measurements of nuclear transparency ratios. They estimated in-medium widths to be in the range of 22–100 MeV \cite{CLAS:2010pxs},   consistent with the measurement from SPring8 \cite{Ishikawa:2004id}.
The ANKE-COSY collaboration \cite{Polyanskiy:2010tj} measured $\phi$ meson production from proton-induced reactions on various nuclear targets and suggested an in-medium $\phi$ width of approximately 50 MeV.
Many theoretical studies have predicted a downward shift in the in-medium $\phi$ meson mass along with a broadening of its decay width. Hatsuda and Lee used the QCD sum rule approach to calculate the in-medium $\phi$ meson mass and predicted a decrease in the range of 1.5\%–3\% at normal nuclear matter density \cite{Hatsuda:1991ez,Hatsuda:1996xt}. In Ref. \cite{Oset:2000eg}, authors, taking into account the renormalization of the \(K\) and \(\bar{K}\) mesons through a coupled channel approach in the medium, 
 forecast a decay width of 22 MeV for $\phi$ mesons. Meanwhile, the work of Ref. \cite{Klingl:1997tm} indicates a negative mass shift of less than 1\% and a decay width of 45 MeV at normal nuclear matter density. In  Ref. \cite{Gubler:2015yna}, authors combined a model based on chiral effective field theory and finite-energy QCD sum rules and observed a downward mass shift of less than 2\% and a significant widening of the width to 45 MeV. Furthermore, in Ref. \cite{Cabrera:2016rnc}, building upon previous work \cite{Hatsuda:1991ez, Hatsuda:1996xt}, the authors reported a negative mass shift of 3.4\% and a substantial decay width of 70 MeV at normal nuclear matter density. 


As stated before, in the current work we shall use an effective chiral SU(3) hadronic mean field model
\cite{Papazoglou:1998vr} to investigate the in-medium properties of kaons and antikaons as well as vector $\phi$ mesons in a finite volume strange hadronic medium.  
To incorporate the finite volume effects, we shall follow the approach of the MRE method, incorporating surface and curvature effects, to evaluate the density of states, as discussed in detail in the later part of the manuscript.
This article is organized as follows. In Sec. \ref{method2}, we briefly describe the framework of the chiral SU(3) model, incorporating corrections for finite-size effects. 
The effective masses of the $K$ and $\Bar{K}$ mesons and the $\phi$ meson mass and decay width in the finite volume hadronic medium are computed in Secs.
\ref{method3} and  \ref{method4}, respectively. Results of present investigations are discussed in Sec. \ref{results}.
Finally, a summary is presented in Sec. \ref{summary}.

\section{Methodology}

\subsection{Hadronic Chiral SU(3) Model} \label{method2}
To study the impact of finite volume on the in-medium masses of
kaons and antikaons,  the chiral SU(3) hadronic mean-field model based on
the nonlinear realization of chiral symmetry 
is used in the present work \cite{Papazoglou:1998vr,Weinberg:1968de,Coleman:1969sm,Bardeen:1969ra}.
In this model, the interactions between baryons are described via the exchange of the scalar fields
$\sigma, \zeta$ and $\delta$ and vector fields $\omega, \rho$ and $\phi$.
The non-strange scalar meson,  $\sigma$, is composed of light $u$ and $d$ quarks, while the strange scalar meson $\zeta$ contains strange $s$ quark \cite{Zakout:1999qu}. The scalar iso-vector field $\delta$ and vector iso-vector field $\rho$ contribute when the medium has finite isospin asymmetry.
Furthermore, the scalar dilaton field $\chi$  representing the glueball field is introduced in the model to incorporate the broken scale invariance property of QCD \cite{Papazoglou:1998vr,Wang:2001hw}.


The Lagrangian density for hadronic chiral SU(3) effective mean-field model is expressed as \cite{Papazoglou:1998vr}

\begin{align}
	\mathcal{L}_{\text{chiral}} &= \mathcal{L}_{\text{kin}} + \sum_{M=S,V,A, Y} \mathcal{L}_{\text{BM}} + \mathcal{L}_{\text{vec}} + \mathcal{L}_{0} + \mathcal{L}_{\text{SB}}. \label{EQ1}
\end{align}


In Eq.(\ref{EQ1}), \(\mathcal{L}_{\text{kin}}\) represents the kinetic energy term for baryons and mesons
and is given by
\begin{align}
	\mathcal{L}_{\text {kin }}&=i \operatorname{Tr} \bar{B} \gamma_{\mu} D^{\mu} B+\frac{1}{2} \operatorname{Tr} D_{\mu} S D^{\mu} S+\operatorname{Tr}\left(u_{\mu} S u^{\mu} S+S u_{\mu} u^{\mu} S\right)
	\quad+\frac{1}{2} \operatorname{Tr} D_{\mu} Y D^{\mu} Y \nonumber \\
	+&\frac{1}{2} D_{\mu} \chi D^{\mu} \chi-\frac{1}{4} \operatorname{Tr}\left(V_{\mu \nu} V^{\mu \nu}\right)-\frac{1}{4} \operatorname{Tr}\left(\mathcal{A}_{\mu \nu} \mathcal{A}^{\mu \nu}\right).  
	\label{Eq_kinetic_2}
\end{align}
In the above equation, the first term represents the kinetic energy term for the baryon octet, $B$. This term also incorporates the interactions of baryons with pseudoscalar mesons through the covariant derivative $D_{\mu}$  defined as $D_\mu B = \partial_\mu B + i \left[\Gamma_\mu, B\right]$, with
$\Gamma_{\mu}=-\frac{i}{4}\left[u^{\dagger} \partial_{\mu} u-\partial_{\mu} u^{\dagger} u+\right.$ $\left.u \partial_{\mu} u^{\dagger}-\partial_{\mu} u u^{\dagger}\right]$.
Here, $u=\exp \left[\frac{i}{\sigma_{0}} \pi^{a} \lambda^{a} \gamma_{5}\right]$ is the unitary transformation operator through which pseudoscalar mesons enter into the calculations.
As discussed in Sec. \ref{method3}, this kinetic term defined in terms of covariant derivative contributes towards the interactions of kaons and antikaons with the nucleons and hyperons in the strange medium (popularly known as Weinberg Tomozawa term). 
The second and third terms of Eq. (\ref{Eq_kinetic_2})  give the kinetic terms for the scalar and pseudoscalar octet mesons, respectively. The fourth and fifth terms represent the kinetic terms for the pseudoscalar singlet $Y$ and the dilaton field $\chi$, respectively. The last two terms of 
Eq. (\ref{Eq_kinetic_2}) defined in terms of field tensors, $ V_{\mu \nu}$ and
$ \mathcal{A}_{\mu \nu}$, represent the  kinetic terms of spin-1 vector and axial vector mesons, respectively.


In Eq.(\ref{EQ1}), the second term, ${\cal{L}}_{BM}$,  gives the interactions of baryons with scalar and vector mesons, which in
the mean-field approximation  is written as
\begin{align}
	\mathcal{L}_{B M}  =-\sum_{i} \overline{\psi_{i}}\left[g_{\omega i} \gamma_{0} \omega+g_{\rho i} \gamma_{0} \tau_{3} \rho+ g_{\phi i} \gamma_0 \phi+ m_{i}^{*} \right] \psi_{i}. \label{Eq_BM_intern}
\end{align}
In the above, $m_i^{*}$ is the effective mass of baryons defined in terms of scalar fields $\sigma, \zeta$ and $\delta$ and is given by
\begin{align}
	m_i^*=-\left(g_{\sigma i} \sigma+g_{\zeta i} \zeta+g_{\delta i} \tau_3 \delta\right) \label{Eq_of_baryon_mass}.
\end{align}
 In Eq. (\ref{Eq_of_baryon_mass}),  symbols \( g_{\sigma i} \), \( g_{\zeta i} \) and \( g_{\delta i} \) denote the coupling strengths associated with baryons $
(i=n,p, \Lambda, \Sigma, \Xi)
$ interacting with the fields \( \sigma \), \( \zeta \), and \( \delta \).
The term $\mathcal{L}_{\text {vec }}$  of Eq. (\ref{EQ1}) gives self interactions of vector mesons, $\mathcal{L}_{\text {0}}$ gives the self interaction of scalar mesons
$\sigma, \zeta$ and $\delta$ and also the interactions for dilaton field $\chi$. The interaction Lagrangian densities for these terms are given as
\begin{align}
	\mathcal{L}_{\text {vec }} & = \frac{1}{2} \frac{\chi^2}{\chi_0^2}\Big(
	m_{\omega}^{2} \omega^ 2+m_{\rho}^{2} \rho^ 2+m_{\phi}^{2} \phi^ 2
	\Big) +g_4 (\omega^4 +2 \phi^4+6 \omega^2 \rho^2+\rho^4),
	\end{align}
and
\begin{align} 
	\mathcal{L}_{\text {0}} & = 
	- \frac{ 1 }{ 2 } k_0 \chi^2
	(\sigma^2+\zeta^2+\delta^2) + k_1 (\sigma^2+\zeta^2+\delta^2)^2
+ k_2 ( \frac{ \sigma^4}{ 2 } + \frac{\delta^4}{2} + \zeta^4
	 +3 \sigma^2 \delta^2)
	  	\nonumber \\
	  	     &   + k_3 \chi (\sigma^2 - \delta^2) \zeta 
	-k_{4} \chi^{4}-\frac{1}{4} \chi^{4} \ln \frac{\chi^{4}}{\chi_{0}^{4}}+\frac{d }{3} \chi^{4} \ln \frac{\left(\sigma^{2}-\delta^{2}\right) \zeta}{\sigma_{0}^{2} \zeta_{0}}, 
\label{Eq_meson_self_scale1}
\end{align}
respectively.
In Eq. (\ref{Eq_meson_self_scale1}), the last two terms account for scale-breaking effects introduced in the chiral SU(3) model through the dilation field $\chi$.
The  term, $\mathcal{L}_{S B}$, of Eq. (\ref{EQ1})
is the explicit symmetry-breaking term and is given by
\begin{align}
	\mathcal{L}_{S B} & =-\left[m_{\pi}^{2} f_{\pi} \sigma+\left(\sqrt{2} m_{K}^{2} f_{K}-\frac{1}{\sqrt{2}} m_{\pi}^{2} f_{\pi}\right) \zeta\right].
\end{align}

The thermodynamic potential, $\Omega$, of the grand canonical ensemble per unit volume $V$, at a given temperature and chemical potential, is expressed as 
\begin{equation}
	\begin{array}{r}
		\frac{\Omega}{V}=-\frac{T \gamma_i}{(2\pi)^3} \sum_i \int d^3 k\left\{\ln \left(e^{-\beta\left[E_i^*(\mathbf{k})-\mu_i^*\right]} +1\right)\right. \\
		\left.+\ln \left(e^{-\beta\left[E_i^*(\mathbf{k})+\mu_i^*\right]} +1\right)\right\}-\mathcal{L}_{\text {vec }}-\mathcal{L}_0-\mathcal{L}_{\mathrm{SB}}-\mathcal{V}_{v a c}, \label{Thermodynaics_potn}
	\end{array}
\end{equation}
where $E_i^*(\mathbf{k}) = \sqrt{\mathbf{k}^2 + m_i^{*2}}$ and $\mu_i^* = \mu_i  - g_{\omega i} \omega-g_{\rho i} \tau_3 \rho$  $ - g_{\phi i} \phi$ are the single-particle effective energy and effective chemical potential, respectively, at temperature $T$. The spin degeneracy factor is represented by $\gamma_i$ and factor, $\beta = \frac{1}{kT}$. In addition, the vacuum potential energy, $\mathcal{V}_{\text{vac}}$, is subtracted from Eq. (\ref{Thermodynaics_potn}) to achieve vanishing vacuum energy.
The equations of motion for the non-strange scalar $\sigma$, the strange scalar $\zeta$, the scalar iso-vector $\delta$, the vector $\omega$, the vector-isovector $\rho$, strange-vector $\phi$, and the scalar dilaton $\chi$ fields are derived by minimizing the thermodynamic potential and are written as

\begin{align}
	& \frac{\partial(\Omega / V)}{\partial \sigma}=k_{0} \chi^{2} \sigma-2 k_{2}\left(\sigma^{3}+3 \sigma \delta^{2}\right) -2 k_{3} \sigma \zeta \chi -4 k_{1}\left(\sigma^{2} +\delta^{2} +\zeta^{2}\right) \sigma \nonumber \\ 
	& -\frac{d}{3} \chi^{4}\left(\frac{2 \sigma}{\sigma^{2}-\delta^{2}}\right)+\left(\frac{\chi}{\chi_{0}}\right)^{2} m_{\pi}^{2} f_{\pi}-\sum_{i} g_{\sigma i} \rho_{i}^{s}=0, \label{eq_sigma} \\
	& \frac{\partial(\Omega / V)}{\partial \zeta}=k_{0} \chi^{2} \zeta-4 k_{2} \zeta^{3}+k_{3} \chi\left(\delta^{2}-\sigma^{2}\right) - 4 k_{1}\left(\sigma^{2}+\delta^{2}+\zeta^{2}\right) \zeta \nonumber \\
	& -\frac{d}{3} \frac{\chi^{4}}{\zeta}+\left(\frac{\chi}{\chi_{0}}\right)^{2}\left[\sqrt{2} m_{K}^{2} f_{K}-\frac{1}{\sqrt{2}} m_{\pi}^{2} f_{\pi}\right]-\sum_{i} g_{\zeta i} \rho_{i}^{s}=0, \label{eq_zeta}
\end{align}
\begin{align}
	& \frac{\partial(\Omega / V)}{\partial \delta}=k_{0} \chi^{2} \delta +2 k_{3} \chi \delta \zeta -2 k_{2}\left(\delta^{3}+3 \sigma^{2} \delta\right)-4 k_{1}\left(\sigma^{2} +\delta^{2} +\zeta^{2}\right) \delta \nonumber \\
	& +\frac{2}{3} d \chi^{4}\left(\frac{\delta}{\sigma^{2}-\delta^{2}}\right)-\sum_{i} g_{\delta i} \tau_{3} \rho_{i}^{s}=0,
	\label{eq_delta}\\
	&\frac{\partial(\Omega / V)}{\partial \omega}=\left(\frac{\chi}{\chi_{0}}\right)^{2} m_{\omega}^{2} \omega+g_{4}\left(12 \rho^{2} \omega + 4 \omega^{3}\right)-\sum_{i} g_{\omega i} \rho_{i}^{v}=0, \label{eq_omega} \\
	&\frac{\partial(\Omega / V)}{\partial \rho}=\left(\frac{\chi}{\chi_{0}}\right)^{2} m_{\rho}^{2} \rho+g_{4}\left( 12 \omega^{2} \rho + 4 \rho^{3} \right)-\sum_{i} g_{\rho i} \tau_{3} \rho_{i}^{v}=0, \label{eq_rho} \\
	&\frac{\partial(\Omega / V)}{\partial \phi}=\left(\frac{\chi}{\chi_{0}}\right)^{2} m_{\phi}^{2} \phi+8 g_{4} \phi^{3}-\sum_{i} g_{\phi i} \rho_{i}^{v}=0, \label{eq_phi}
\end{align}
and
\begin{align}
	&\frac{\partial(\Omega / V)}{\partial \chi}=k_{0} \chi\left(\sigma^{2}+\delta^{2}+\zeta^{2}\right)+\left(4 k_{4}-d\right) \chi^{3}+k_{3}\left(\delta^{2}-\sigma^{2}\right) \zeta+\chi^{3}\left[1+\ln \left(\frac{\chi^{4}}{\chi_{0}^{4}}\right)\right] \nonumber \\
	&-\frac{4}{3} d \chi^{3} \ln \left(\left(\frac{\left(\sigma^{2}-\delta^{2}\right) \zeta}{\sigma_{0}^{2} \zeta_{0}}\right)\left(\frac{\chi}{\chi_{0}}\right)^{3}\right)+\frac{2 \chi}{\chi_{0}^{2}}\left[m_{\pi}^{2} f_{\pi} \sigma+\left(\sqrt{2} m_{K}^{2} f_{K}-\frac{1}{\sqrt{2}} m_{\pi}^{2} f_{\pi}\right) \zeta\right] \nonumber \\
	&-\frac{\chi}{\chi_{0}^{2}}\left(m_{\omega}^{2} \omega^{2}+m_{\rho}^{2} \rho^{2} + m_{\phi}^{2} \phi^{2}\right)=0, 
	\label{eq_chi}
\end{align}
respectively. The scalar and vector densities of the $i$th baryon appearing in the above equation are defined as
\begin{align}
	\rho_i^s=\gamma_i \int \frac{d^3 k}{(2 \pi)^3} \frac{m_i^*}{E_i^*(\mathbf{k})}\left(n_i(\beta)+\bar{n}_i(\beta)\right),
	\label{Eq_scalar_d}
\end{align}
and
\begin{align}
	\rho_i^v=\gamma_i \int \frac{d^3 k}{(2 \pi)^3}\left(n_i(\beta)-\bar{n}_i(\beta)\right),
	\label{Eq_vector_d}
\end{align}
with
\begin{align}
	n_i(\beta)=\frac{1}{\exp \left[\beta\left(E_i^*(\mathbf{k})-\mu_i^*\right) \right]+1},   \quad \bar{n}_i(\beta)=\frac{1}{\exp \left[\beta\left(E_i^*(\mathbf{k})+\mu_i^*\right)\right] + 1}, \label{Eq_distribution_fun}
\end{align}
as the finite temperature distribution functions of baryons and antibaryons, respectively.
In the present work, we shall investigate the in-medium properties of pseudoscalar and vector mesons at finite isospin asymmetry and finite strangeness fraction of the medium. The isospin asymmetry is introduced through the parameter $\eta = \frac{-\sum_i I_{3i} \rho_i^v}{ \rho_B}$, whereas for the finite strangeness fraction the definition  $f_s = \frac{\sum_i |s_i| \rho_i^v}{\rho_B}$
is used. The symbols $I_{3i}$, $|s_i|$, and $\rho_B$ represent the isospin quantum number (third component), the number of strange quarks, and the total baryonic density, respectively.


As discussed previously, 
in the current work, the medium modification of kaons and antikaons and the vector $\phi$ mesons will
be studied in finite volume strange hadronic matter at finite temperature. 
To achieve this, the coupled
system of equations, given by
Eqs. (\ref{eq_sigma})- 
(\ref{eq_chi}) for the scalar and vector fields are solved, incorporating finite volume corrections through the MRE method 
\cite{Zhao:2018nqa,Kiriyama:2005eh,Balian:1970fw,He:1996fc}.
Under the MRE method, 
the expression for the density of states, for a medium with finite volume and spherical geometry, having radius $R$, is written as \cite{Kiriyama:2002xy,MataCarrizal:2022gdr,Lugones:2013ema}
\begin{equation}
	\rho_{M R E}=\frac{k^{2}}{2 \pi^{2}}\left[1+\frac{6 \pi^{2}}{k R} f_{S}\left(\frac{k}{m_i^*}\right)+\frac{12 \pi^{2}}{(k R)^{2}} f_{C}\left(\frac{k}{m_i^*}\right)\right]. \label{Eq_MRE_gen}
\end{equation}
In the above equation, $f_S$ and $f_C$ are the surface and curvature contributions to the density of states and are given by
\begin{align}
	f_{S}\left(\frac{k}{m_i^*}\right)=-\frac{1}{8 \pi}\left[1-\frac{2}{\pi} \arctan \frac{k}{m_i^*}\right], \label{Eq_surface_term_mre}
\end{align}
and
\begin{align}
	f_{C}\left(\frac{k}{m_i^*}\right)=\frac{1}{12 \pi^{2}}\left[1-\frac{3 k}{2 m_i^*}\left(\frac{\pi}{2}-\arctan \frac{k}{m_i^*}\right)\right], \label{Eq_curvature_term_mre}
\end{align}
respectively.
Also, as defined earlier, $m_i^*$   is the effective mass of baryons in the finite volume hadronic medium.
In literature, the finite volume effects with surface and curvature contributions have been investigated for quark matter only,  with 
$m_i^*$ being the constituent quark mass \cite{Kiriyama:2005eh,Kiriyama:2002xy,Kiriyama:2006uh}.
In some of these studies,  $m_i^*$ is being replaced with the parameter  $\alpha$ defining the penetration of wave function from finite volume quark matter to hadronic phase
and its values are taken as, $\alpha = \infty$, enforcing Dirichlet boundary conditions  and $\alpha = 0$ in the case when Neumann boundary conditions are implemented \cite{Balian:1970fw}.
For investigating the properties of finite volume hadronic medium in the present work, we restrict to these two values for $\alpha$, i.e., $\alpha = \infty$ and $0$.

\begin{figure}[htbp]
	\centering
	\includegraphics[width=0.6\textwidth]{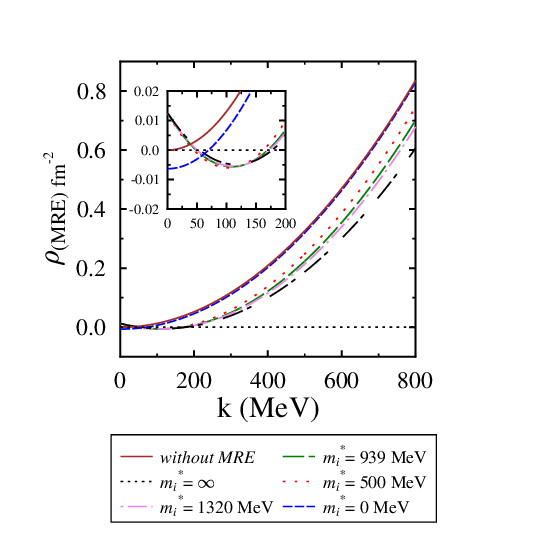}
	\caption{In the above figure, the variation of density of states, $\rho_{MRE}$, as a function of momentum, $k$, is shown for various values effective masses of baryons, $m_i^*$, in the finite volume hadronic medium. Results are compared with the case when the density of states is calculated without the MRE method. \label{plot_of_density}}
\end{figure}

As is apparent from 
Eq. (\ref{Eq_MRE_gen}), due to quadratic nature in $k$,
the density of the state has two roots.
Between these roots, there can be  density of states that have negative values and, thus, have no physical significance. An infrared (IR) cut-off, $\Lambda_{IR}$, is introduced in the volume integral over momentum states appearing in the definitions of scalar and vector densities of baryons defined through Eqs. (\ref{Eq_scalar_d}) and (\ref{Eq_vector_d}), to eliminate   negative values of density of states  through the following prescription  \cite{Kiriyama:2005eh,Zhao:2018nqa}
\begin{align}
	\int_0^{ \infty} \frac{d^3 k}{(2 \pi)^3} \cdots \rightarrow \int_{\Lambda_{ I R}}^{ \infty} \frac{d^3 k}{(2 \pi)^3} \rho_{ M R E} \cdots,
	\label{Eq_den_MREint}
\end{align}
where $\Lambda_{I R}$ is the largest solution to, $\rho_{M R E}(k)=0$, with respect to momentum $k$.
With  $\alpha = \infty $, the Dirichlet boundary conditions, the density of states will become \cite{Madsen:1994vp,Kiriyama:2006uh}
\begin{equation}
	\rho_{M R E}=\frac{k^{2}}{2 \pi^{2}}\left[1-\frac{3 \pi}{4 k R}+\frac{1}{(k R)^{2}}\right] ,
\end{equation}
with $\Lambda_{I R}=1.8 / R$. 
On the other side, implementing the Neumann boundary conditions, i.e., $\alpha = 0$, the expression for the density of states is given by 
\begin{align}
	\rho_{M R E}=\frac{k^{2}}{2 \pi^{2}}\left[1-\frac{1}{2(k R)^{2}}\right], \label{rho_mre_NBC}
\end{align}
which leads to $\Lambda_{I R}=1 / \sqrt{2} R$.
The above discussion  for the density of states in the MRE method consider the finite volume in a spherical geometry, which takes into account surface and curvature effects. As outlined in Ref. \cite{MataCarrizal:2022gdr},  considering cubic geometry for finite volume matter, the density of states in the MRE method
can be written as 
%
\begin{equation}
	\rho_{M R E}=\frac{k^{2}}{2 \pi^{2}}\left(1-\frac{3}{4 k R}\right), 
	\label{eq_cubic_L}
\end{equation}
with IR cut-off conditions $\Lambda_{I R}=0.75 / R$, where now $R$ being the length of cubic volume.  
For a better understanding of above discussed different 
cases,  in Fig. (\ref{plot_of_density})
densities of states, $\rho_{M R E}$, calculated using Eq. (\ref{Eq_MRE_gen})
is plotted
 as a function of momentum, $k$, for  various values of effective mass, $m_i^*$, of baryons  in the hadronic medium with finite volume.
 All results are plotted for system size $R = 2$ fm and compared with the situation when densities of states are calculated without using the MRE method, in which case they do not become negative for any value of $k$.
As can be seen from this figure, the value of $k$ at which the density of states become negative decrease with a decrease in $m_i^*$.
Thus, in the Dirichlet case, i.e., when $ \alpha = m_i^* = \infty$, the value of $k$ and hence, the lower cut-off $\Lambda_{ I R}$ will have largest value compared to other cases and maximum states are excluded (as they have negative value) from the integral in Eq. (\ref{Eq_den_MREint}).   
\subsection{Kaon and antikaon in the hadronic chiral SU(3) model}
 \label{method3}
%
To obtain the medium modified masses of kaons and antikaons in the finite volume strange hadronic matter within the chiral SU(3) mean field model, the  Lagrangian density
describing the interactions
of kaons and antikaons with the baryon octet and the scalar mesons $\sigma, \zeta$ and $\delta$ is written as \cite{Mishra:2008dj}
\begin{align}
	\mathcal{L}_{K B}= & -\frac{i}{4 f_{K}^{2}} \left[ \left( 2 \bar{p} \gamma^{\mu} p + \bar{n} \gamma^{\mu} n 
	- \overline{\Sigma^{-}} \gamma^{\mu} \Sigma^{-} + \overline{\Sigma^{+}} \gamma^{\mu} \Sigma^{+} 
	- 2 \bar{\Xi}^{-} \gamma^{\mu} \Xi^{-} - \overline{\Xi^{0}} \gamma^{\mu} \Xi^{0} \right) \right. \notag \\
	& \quad \times \left( K^{-}\left(\partial_{\mu} K^{+}\right) - \left(\partial_{\mu} K^{-}\right) K^{+} \right) \notag \\
	& \quad + \left( \bar{p} \gamma^{\mu} p + 2 \bar{n} \gamma^{\mu} n + \overline{\Sigma^{-}} \gamma^{\mu} \Sigma^{-} 
	- \overline{\Sigma^{+}} \gamma^{\mu} \Sigma^{+} - \overline{\Xi^{-}} \gamma^{\mu} \Xi^{-} 
	- 2 \overline{\Xi^{0}} \gamma^{\mu} \Xi^{0} \right) \notag \\
	& \quad \times \left( \overline{K^{0}}\left(\partial_{\mu} K^{0}\right) 
	- \left(\partial_{\mu} \overline{K^{0}}\right) K^{0} \right) \bigg] \notag \\
	& + \frac{m_{K}^{2}}{2 f_{K}} \left[ (\sigma + \sqrt{2} \zeta + \delta)\left(K^{+} K^{-}\right) 
	+ (\sigma + \sqrt{2} \zeta - \delta)\left(K^{0} \bar{K}^{0}\right) \right] \notag \\
	& - \frac{1}{f_{K}} \left[ (\sigma + \sqrt{2} \zeta + \delta)\left(\partial_{\mu} K^{+}\right)\left(\partial^{\mu} K^{-}\right) 
	+ (\sigma + \sqrt{2} \zeta - \delta)\left(\partial_{\mu} K^{0}\right)\left(\partial^{\mu} \overline{K^{0}}\right) \right] \notag \\
	& + \frac{d_{1}}{2 f_{K}^{2}} \left( \bar{p} p + \bar{n} n + \bar{\Lambda}^{0} \Lambda^{0} + \bar{\Sigma}^{-} \Sigma^{+} 
	+ \bar{\Sigma}^{0} \Sigma^{0} + \bar{\Sigma}^{-} \Sigma^{-} + \bar{\Xi}^{-} \Xi^{-} + \bar{\Xi}^{0} \Xi^{0} \right) \notag \\
	& \quad \times \left( \left(\partial_{\mu} K^{+}\right)\left(\partial^{\mu} K^{-}\right) 
	+ \left(\partial_{\mu} K^{0}\right)\left(\partial^{\mu} \bar{K}^{0}\right) \right) \notag \\
	& + \frac{d_{2}}{2 f_{K}^{2}} \left[ \left( \bar{p} p + \frac{5}{6} \Lambda^{0} \Lambda^{0} + \frac{1}{2} \bar{\Sigma}^{0} \Sigma^{0} 
	+ \bar{\Sigma}^{+} \Sigma^{+} + \bar{\Xi}^{-} \Xi^{-} + \bar{\Xi}^{0} \Xi^{0} \right) \right. \notag \\
	& \quad \times \left( \partial_{\mu} K^{+} \right) \left( \partial^{\mu} K^{-} \right) \notag \\
	& \quad + \left( \bar{n} n + \frac{5}{6} \bar{\Lambda}^{0} \Lambda^{0} + \frac{1}{2} \bar{\Sigma}^{0} \Sigma^{0} 
	+ \overline{\Sigma^{-}} \Sigma^{-} + \bar{\Xi}^{-} \Xi^{-} + \bar{\Xi}^{0} \Xi^{0} \right) \notag \\
	& \quad \times \left( \partial_{\mu} K^{0} \right) \left( \partial^{\mu} \bar{K}^{0} \right) \bigg]. \label{Eq_interaction_KB}
\end{align}

In the above equation,  $K^+$ and  $K^0$ belong to
kaon, $K$, isospin doublet, whereas,   $K^-$ and $\bar{K}^0$ are the members of antikaon $\bar{K}$ doublet.
In Eq. (\ref{Eq_interaction_KB}),  the first term (first four lines in the square bracket) is the vectorial  Weinberg Tomozawa interaction term, obtained from the first term of Eq. (\ref{Eq_kinetic_2}).
The second term, which generates an attractive interaction for kaons and antikaons as a function of the density of the medium, arises from the explicit symmetry-breaking term of the model \cite{Papazoglou:1998vr,Mishra:2006wy}.
In the framework of the chiral SU(3) model, the third term
of
Eq. (\ref{Eq_interaction_KB}) is derived from the kinetic part of the pseudoscalar mesons (third term of Eq. (\ref{Eq_kinetic_2}))  \cite{Mishra:2006wy,Mishra:2008kg}. Fourth and fifth terms in Eq. (\ref{Eq_interaction_KB}), named as range terms, 
arise from the Lagrangian densities \cite{Mishra:2004te,Mishra:2006wy}
\begin{equation}
	\mathcal{L}_{\left(d_{1}\right)}^{B M}=d_{1} \operatorname{Tr}\left(u_{\mu} u^{\mu} \bar{B} B\right),
\end{equation}
and                             
\begin{equation}
	\mathcal{L}_{\left(d_{2}\right)}^{B M}=d_{2} \operatorname{Tr}\left(\bar{B} u_{\mu} u^{\mu} B\right).
\end{equation}
 From the interaction  Lagrangian density given by Eq.(\ref{Eq_interaction_KB}),
equation of motion for kaons and antikaons are derived using Euler's Lagrange equation of motion, whose
Fourier transformation   gives the dispersion relation,
\begin{equation}
	\vec{k}^{2} -\omega^{2} +m_{K(\Bar{K})}^{2}=\Pi_{K(\Bar{K})}^{*}(\omega,|\vec{k}|, \rho), \label{Eq_dispertion}
\end{equation}

where $\Pi_{K(\Bar{K})}^{*}(\omega, |\vec{k}|, \rho)$ denotes the kaon (antikaon) self-energy in the medium.
Also, $m_{K(\Bar{K})}$ in Eq. (\ref{Eq_dispertion}) represents the vacuum mass of kaon (antikaon). 
Specifically,   for the kaon doublet, $K$ $(K^+, {K^0})$,  self energy is given as
\begin{align}
	\Pi_{K}^{*}(\omega,|\vec{k}|, \rho) = & -\frac{1}{4 f_{K}^{2}} \left[ 
	3\left(\rho_{p}^{v} + \rho_{n}^{v}\right) \pm \left(\rho_{p}^{v} - \rho_{n}^{v}\right) 
	\pm 2\left(\rho_{\Sigma^{+}}^{v} - \rho_{\Sigma^{-}}^{v}\right) 
	- 3\left(\rho_{\Xi^{-}}^{v} + \rho_{\Xi^{0}}^{v}\right) 
	\pm \left(\rho_{\Xi^{-}}^{v} - \rho_{\Xi^{0}}^{v}\right) 
	\right] \omega \nonumber \\
	& + \frac{m_{K}^{2}}{2 f_{K}} \left( \sigma^{\prime} + \sqrt{2} \zeta^{\prime} \pm \delta^{\prime} \right) \nonumber \\
	& + \left[
	-\frac{1}{f_{K}} \left( \sigma^{\prime} + \sqrt{2} \zeta^{\prime} \pm \delta^{\prime} \right) 
	+ \frac{d_{1}}{2 f_{K}^{2}} \left( \rho_{p}^{s} + \rho_{n}^{s} + \rho_{\Lambda^{0}}^{s} + \rho_{\Sigma^{+}}^{s} 
	+ \rho^{s}_{\Sigma^{0}} + \rho_{\Sigma^{-}}^{s} + \rho^{s}_{\Xi^{-}} + \rho^{s}_{\Xi^{0}} \right)
	\right. \nonumber \\
	& \quad + \frac{d_{2}}{4 f_{K}^{2}} \left( 
	\left( \rho^{s}_{p} + \rho^{s}_{n} \right) \pm \left( \rho^{s}_{p} - \rho^{s}_{n} \right) 
	+ \rho^{s}_{\Sigma^{0}} + \frac{5}{3} \rho^{s}_{\Lambda^{0}} 
	+ \left( \rho^{s}_{\Sigma^{+}} + \rho^{s}_{\Sigma^{-}} \right) \pm \left( \rho_{\Sigma^{+}}^{s} - \rho_{\Sigma^{-}}^{s} \right)
	\right.\nonumber\\
	& \quad \left. \left. 
	+ 2 \rho_{\Xi^{-}}^{s} + 2 \rho_{\Xi^{0}}^{s}
	\right) \right] \left( \omega^{2} - \vec{k}^{2} \right),  \label{Eq_dis_kaon}
\end{align}

where contributions to $K^{+}$ and $K^{0}$ are denoted by the $\pm$ signs, respectively. Also, $\sigma^{\prime}$ $=\sigma-\sigma_{0}$, $\zeta^{\prime}$ $=\zeta-\zeta_{0}$, and $\delta^{\prime}$ $=\delta-\delta_{0}$ are the variation of $\sigma$, $\zeta$ and $\delta$ fields from their expectation values in vacuum. A non-zero value for $\delta$ will break the isospin symmetry in the medium (in vacuum, $\delta_{0}=0$). 

Likewise, the expression for the self-energy of antikaon isospin doublet, $\bar{K}$ $(K^-, \bar{K^0})$, is written as
\begin{align}
	\Pi_{\bar{K}}^{*}(\omega,|\vec{k}|, \rho) = & \frac{1}{4 f_{K}^{2}} \left[ 
	3\left(\rho_{p}^{v} + \rho_{n}^{v}\right) \pm \left(\rho_{p}^{v} - \rho_{n}^{v}\right) 
	\pm 2\left(\rho_{\Sigma^{+}}^{v} - \rho_{\Sigma^{-}}^{v}\right) 
	- 3\left(\rho_{\Xi^{-}}^{v} + \rho_{\Xi^{0}}^{v}\right) 
	\pm \left(\rho_{\Xi^{-}}^{v} - \rho_{\Xi^{0}}^{v}\right) 
	\right] \omega \nonumber \\
	& + \frac{m_{K}^{2}}{2 f_{K}} \left( \sigma^{\prime} + \sqrt{2} \zeta^{\prime} \pm \delta^{\prime} \right) \nonumber \\
	& + \left[
	-\frac{1}{f_{K}} \left( \sigma^{\prime} + \sqrt{2} \zeta^{\prime} \pm \delta^{\prime} \right) 
	+ \frac{d_{1}}{2 f_{K}^{2}} \left( \rho^{s}_{p} + \rho^{s}_{n} + \rho_{\Lambda^{0}}^{s} + \rho_{\Sigma^{+}}^{s} 
	+ \rho^{s}_{\Sigma^{0}} + \rho_{\Sigma^{-}}^{s} + \rho^{s}_{\Xi^{-}} + \rho^{s}_{\Xi^{0}} \right)
	\right. \nonumber \\
	& \quad + \frac{d_{2}}{4 f_{K}^{2}} \left( 
	\left( \rho_{p}^{s} + \rho^{s}_{n} \right) \pm \left( \rho_{p}^{s} - \rho^{s}_{n} \right) 
	+ \rho^{s}_{\Sigma^{0}} + \frac{5}{3} \rho^{s}_{\Lambda^{0}} 
	+ \left( \rho^{s}_{\Sigma^{+}} + \rho^{s}_{\Sigma^{-}} \right) \pm \left( \rho_{\Sigma^{+}}^{s} - \rho_{\Sigma^{-}}^{s} \right)
	\right. \nonumber \\
	& \quad \left. \left. 
	+ 2 \rho_{\Xi^{-}}^{s} + 2 \rho_{\Xi^{0}}^{s}
	\right) \right] \left( \omega^{2} - \vec{k}^{2} \right),  \label{Eq_dis_antikaon}
\end{align}

where  $K^{-}$and $\bar{K}^{0}$ are denoted by the $\pm$ signs, respectively.
The effective mass of the \(K\) (\(\bar{K}\)) meson in the strange hadronic medium with finite volume is determined by solving Eq. (\ref{Eq_dispertion}) with the constraint \(m^*_{K(\bar{K})} = \omega(|\mathbf{k}| = 0)\).
The finite volume effects in the in-medium masses of kaons and antikaons are introduced through the modification of various quantities  appearing in the self-energy terms, which are calculated in the finite  volume hadronic medium as described in the previous section, i.e,  in terms of
scalar fields $\sigma, \zeta$, and $\delta$,  the scalar
densities $\rho_i^{s}$ and vector densities $\rho_i^v$ of the baryons.
The values of parameters \(d_1\) and \(d_2\) used in the
expressions of self-energy are chosen as ${2.56}/{m_K}$ and 
${0.73}/{m_K}$ respectively, fitted using empirical data on $KN$ scattering length \cite{Mishra:2006wy,Barnes:1992ca}.

\subsection{$\phi$ meson self energy} \label{method4}

\begin{figure}[h]
	\centering
	\includegraphics[width=0.55\textwidth]{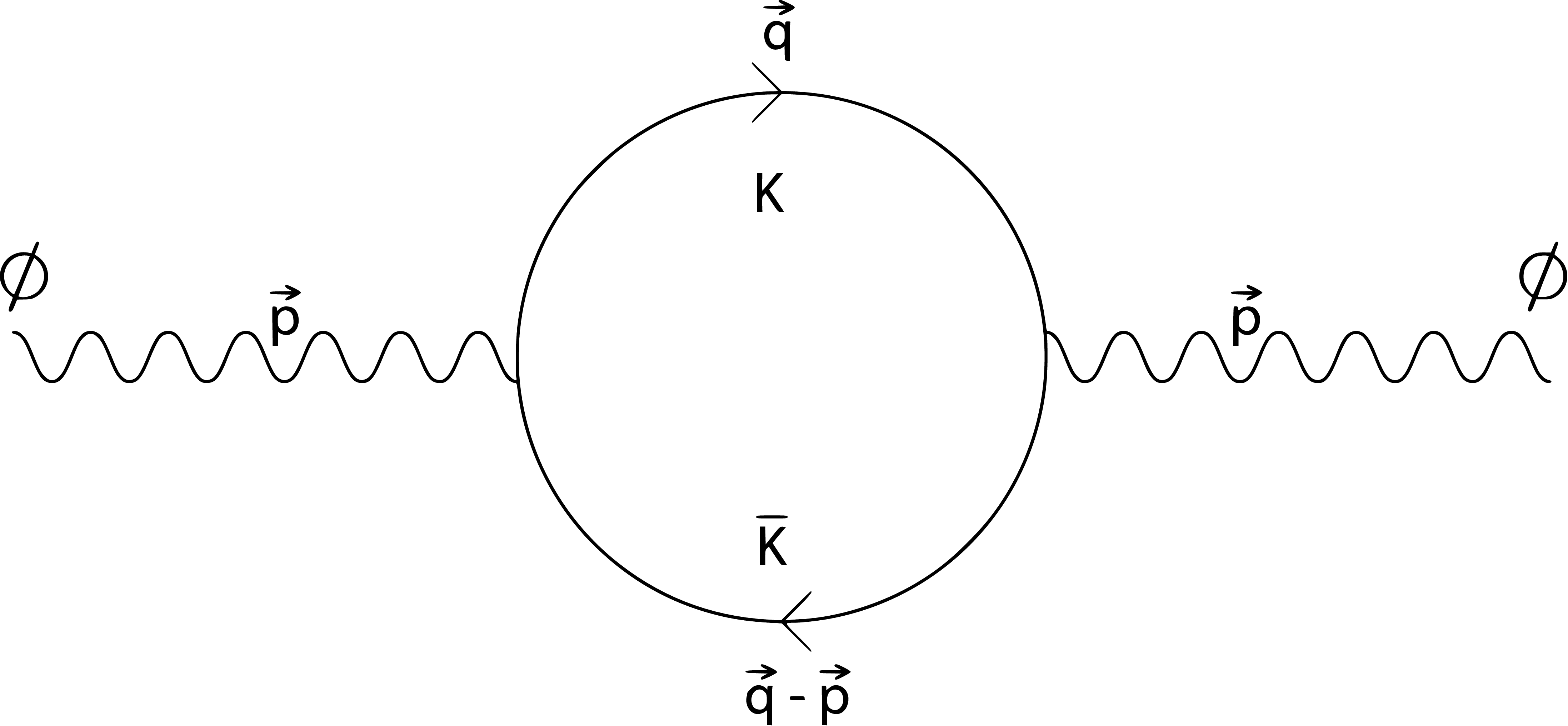}
	\caption{The one loop level interaction of $\phi K \bar{K}$ \cite{Cobos-Martinez:2017vtr}.\label{fig:1}}
\end{figure}

The effective masses of kaons and antikaons calculated in the finite volume hadronic medium are used as input in the present section to calculate the in-medium masses and decay width of $\phi$ mesons. To calculate the $\phi$ meson self-energy
in the hadronic medium, we consider  the decay process $\phi \rightarrow K \bar{K}$ at the one-loop (as shown in Figure \ref{fig:1}), for which the interaction Lagrangian  density is written as  \cite{Cobos-Martinez:2017vtr,Ko:1992tp} 
\begin{equation}
	\mathcal{L}_{\text {int }}= 
	i g_\phi \phi^\mu\left[\bar{K}\left(\partial_\mu K\right)-\left(\partial_\mu \bar{K}\right) K\right]. 
\end{equation}
In the above equation, $g_{\phi}$ represents the coupling constant.
%
Since the contributions of $\phi \phi$K$\Bar{K}$ interactions to the in-medium masses and decay width are smaller than those of $\phi$K$\Bar{K}$ interactions, hence these interactions are not considered in the current work \cite{Cobos-Martinez:2017vtr}. In the rest frame of $\phi$ mesons, the scalar component of the self-energy, corresponding to the loop diagram as shown in Fig. (\ref{fig:1}),
 can be expressed as  
\begin{equation}
	i \Pi_{\phi}^{*}(p)=-\frac{8}{3} g_{\phi}^{2} \int \frac{d^{4} q}{(2 \pi)^{4}} \vec{q }^{2} D_{K}(q) D_{\bar{K}}(q-p), \label{32}
\end{equation}
where, $D_{K}(q)=\left(-m_{K}^{*^{2}} + q^{2} +i \epsilon\right)^{-1}$ and $D_{\bar{K}}(q-p)=\left(-m_{\bar{K}}^{*^{2}} + (q-p)^{2} +i \epsilon\right)^{-1}$ represent the propagators for kaons and antikaons, respectively. Here, $p=\left(p^{0}=m_{\phi}^{*}, \overrightarrow{0}\right)$ denotes the four-momentum vector of the $\phi$ meson, with $m_{\phi}^{*}$ as the in-medium   $\phi$ meson mass. Also, $m_{K}^{*}\left(=\frac{m_{K^{0}}^{*} + m_{K^{+}}^{*}}{2}\right)$ and $m_{\bar{K}}^{*}\left(=\frac{m_{\Bar{K}^{0}}^{*}+m_{K^{-}}^{*}}{2}\right)$, are the averaged kaons and antikaon masses, respectively. The values of $m_{K^{+}}^{*}, m_{K^{0}}^{*}, m_{K^{-}}^{*}$, and $m_{\bar{K}^{0}}^{*}$ are computed using Eq. (\ref{Eq_dispertion}) in  the finite volume strange hadronic medium, at finite temperature and for finite isospin asymmetry. The in-medium mass of the $\phi$ meson is calculated from the real part of $\Pi_{\phi}^{*}(p)$, as determined by the following relation
\begin{equation}
	m_{\phi}^{*^{2}}=\left(m_{\phi}^{0}\right)^{2}+\operatorname{Re} \Pi_{\phi}^{*}\left(m_{\phi}^{*^{2}}\right), \label{33}
\end{equation}
where $m_{\phi}^{0}$ is the bare mass of the $\phi$ meson. The real part of the self-energy can be written \cite{Cobos-Martinez:2017vtr,Kumar:2020vys} as 
\begin{equation}
	\operatorname{Re} \Pi_{\phi}^{*}=-\frac{4}{3} g_{\phi}^{2} \mathcal{P} \int \frac{d^{3} q}{(2 \pi)^{3}} \vec{q}^{2} \frac{\left(E_{K}^{*}+E_{\bar{K}}^{*} v\right)}{E_{K}^{*} E_{\bar{K}}^{*}\left(\left(E_{K}^{*}+E_{\bar{K}}^{*}\right)^{2}-m_{\phi}^{*^{2}}\right)}, \label{34}
\end{equation}
where $\mathcal{P}$ denotes the principal value of the integral given in Eq. (\ref{32}). Also, $E_{K}^{*}=$ $\left(\vec{q}^{\,2}+m_{K}^{*^{2}}\right)^{1 / 2}$, and $E_{\bar{K}}^{*}=\left(\vec{q}^{\,2}+m_{\bar{K}}^{*^{2}}\right)^{1 / 2}$.
To circumvent singularities, the integral in Eq. (\ref{34}) is regularized by incorporating a phenomenological form factor with the cutoff parameter $\Lambda_{c}$, i.e., \cite{Krein:2010vp} 

\begin{align}
	\operatorname{Re} \Pi_{\phi}^{*}=  -\frac{4}{3} g_{\phi}^{2} \mathcal{P} \int_{0}^{\Lambda_{c}} \frac{d^{3} q}{(2 \pi)^{3}} \vec{q}^{2}\left(\frac{\Lambda_{c}^{2}+m_{\phi}^{*^{2}}}{\Lambda_{c}^{2}+4 E_{K}^{*^{2}}}\right)^{4} 
	\frac{\left(E_{K}^{*}+E_{\bar{K}}^{*}\right)}{E_{K}^{*} E_{\bar{K}}^{*}\left(\left(E_{K}^{*}+E_{\bar{K}}^{*}\right)^{2}-m_{\phi}^{*^{2}}\right)}. \label{real_part_of_self_eng_phi}
\end{align}
 The decay width of the $\phi$ meson is determined by extracting the imaginary component of its self-energy, denoted as $\operatorname{Im} \Pi_{\phi}^{*}$, and is expressed in terms of  in-medium masses of  $\phi$, $K$, and $\Bar{K}$ mesons \cite{Li:1994cj}, i.e.,
\begin{align}
	\Gamma_{\phi}^{*}=  \frac{g_{\phi}^{2}}{24 \pi} \frac{1}{m_{\phi}^{*^{5}}}\left[\left(m_{\phi}^{*^{2}}-\left(m_{K}^{*}+m_{\bar{K}}^{*}\right)^{2}\right)\right.
	\left.\times\left(m_{\phi}^{*^{2}}-\left(m_{K}^{*}-m_{\bar{K}}^{*}\right)^{2}\right)\right]^{3 / 2}. \label{phi_decay_width}
\end{align}
In the current study, the value of coupling constant $g_{\phi}$ is taken to be $4.539$, fitted to the empirical width of the $\phi$ meson in vacuum. 

\section{Results and discussion}  \label{results}

In this section, we present the results  on the in-medium masses of kaons and antikaons alongwith the masses and decay width of $\phi$ mesons in a strange hadronic medium with finite volume. 
As discussed earlier, within the chiral SU(3) model, the in-medium masses of baryons are calculated through the
medium modified scalar fields $\sigma, \zeta$ and $\delta$, which further modify the properties of kaons and antikaons.
In  Sec. \ref{method5}, we shall discuss the impact of finite volume isospin asymmetric strange hadronic medium  at finite temperature on the scalar fields  $\sigma, \zeta$ and $\delta$
as well as the dilaton field $\chi$. The in-medium properties of kaons and antikaons are presented in Sec. \ref{method6}, and finally, Sec. \ref{method7} is dedicated to the
discussion on the results for the 
in-medium masses and decay widths of $\phi$ mesons. The parameters  of the chiral SU(3) model used in the present study are listed in table
\ref{tab:Table 1}.

\begin{table}[htbp]
    \centering
    \caption{Various parameters used in the present calculations \cite{Kumar:2011ff}}
    \label{tab:Table 1}
    \begin{tabular}{|c|c|c|c|c|c|c|c|c|c|}
        \hline
        $g_{\sigma n(p)}$ & $g_{\zeta n(p)}$ & $g_{\delta n(p)}$ & $g_{\omega n(p)}$ & $g_{\rho n(p)}$ & $g_{\sigma \Lambda}$ & $g_{\zeta \Lambda}$ & $g_{\delta \Lambda}$ & $g_{\sigma \Sigma}$ & $g_{\zeta \Sigma}$ \\
        \hline
        10.6 & -0.46 & 2.48 & 13.32 & 5.48 & 5.31 & 5.8 & 0 & 6.13 & 5.8 \\
        \hline
        $g_{\delta \Sigma}$ & $g_{\delta \Sigma^0}$ & $g_{\sigma \Xi}$ & $g_{\zeta \Xi}$ & $g_{\delta \Xi}$ & $g_{\omega \Lambda}$ & $g_{\omega \Sigma}$ & $g_{\rho \Sigma}$ & $g_{\rho \Sigma^0}$ & $g_{\omega \Xi}$ \\
        \hline
        6.79 & 0 & 3.68 & 9.14 & 2.36 & $\frac{2}{3} g_{\omega N}$ & $\frac{2}{3} g_{\omega N}$ & $\frac{2}{3} g_{\omega N}$ & 0 & $\frac{1}{3} g_{\omega N}$ \\
        \hline
        $g_{\rho \Lambda}$ & $g_{\rho \Xi}$ & $g_{\phi \Lambda}$ & $g_{\phi \Sigma}$ & $g_{\phi \Xi}$ & $\sigma_0(\mathrm{MeV})$ & $\zeta_0(\mathrm{MeV})$ & $\chi_0(\mathrm{MeV})$ & $d$ & $\rho_0\left(\mathrm{fm}^{-3}\right)$ \\
        \hline
        0 & $\frac{1}{3} g_{\omega N}$ & $-\frac{\sqrt{2}}{3} g_{\omega N}$ & $-\frac{\sqrt{2}}{3} g_{\omega N}$ & $-\frac{2 \sqrt{2}}{3} g_{\omega N}$ & -93.29 & -106.8 & 409.7 & 0.064 & 0.15 \\
        \hline
        $m_\pi(\mathrm{MeV})$ & $m_K(\mathrm{MeV})$ & $f_\pi(\mathrm{MeV})$ & $f_K(\mathrm{MeV})$ & $g_4$ & $k_0$ & $k_1$ & $k_2$ & $k_3$ & $k_4$ \\
        \hline
        139 & 498 & 93.29 & 122.14 & 79.91 & 2.53 & 1.35 & -4.77 & -2.77 & -0.218 \\
        \hline
        $m_\omega(\mathrm{MeV})$ & $m_\rho(\mathrm{MeV})$ & $m_\phi(\mathrm{MeV})$ & & & & & & & \\
        \hline
        780.56 & 761.06 & 1019 & & & & & & & \\
        \hline
    \end{tabular}
\end{table}

\subsection{Scalar fields of the chiral SU(3) hadronic model in finite volume matter} \label{method5}
\begin{figure}[htbp]
\centering
\includegraphics[width=0.75\textwidth]{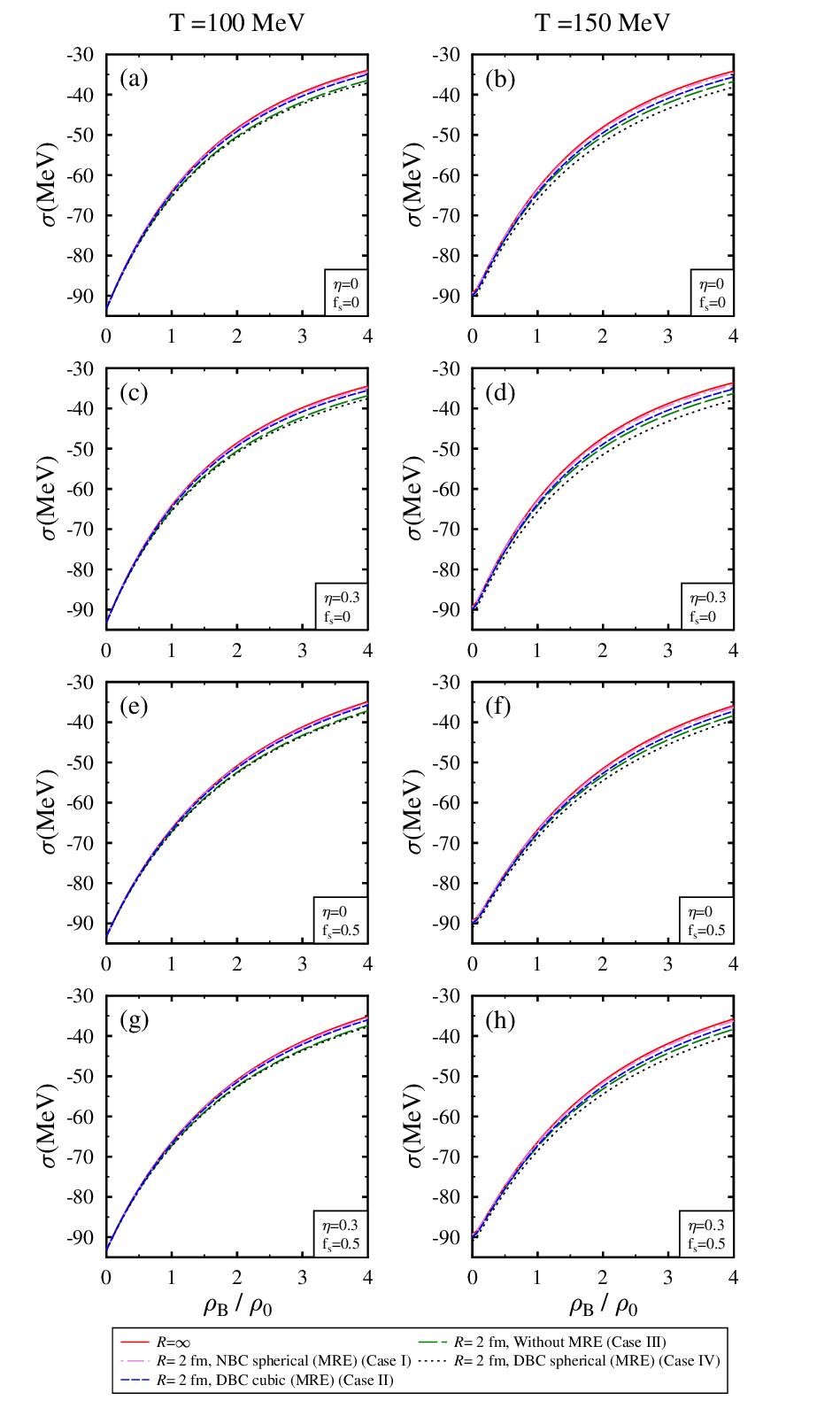}
\caption{ The variation of scalar field $\sigma$ is shown   with respect to the baryon density $\rho_B$ (in units of nuclear saturation density $\rho_0 $),  at  temperatures, $T = 100$  [in subplots (a), (c), (e) and (g)] and $150$ MeV [in subplots (b), (d), (f) and (h)]. 
 In the above, NBC and DBC are for Neumann and Dirichlet boundary conditions, and MRE denotes multiple reflection expansion.  
\label{fig:2}}
\end{figure}

\begin{figure}[htbp]
\centering
\includegraphics[width=0.75\textwidth]{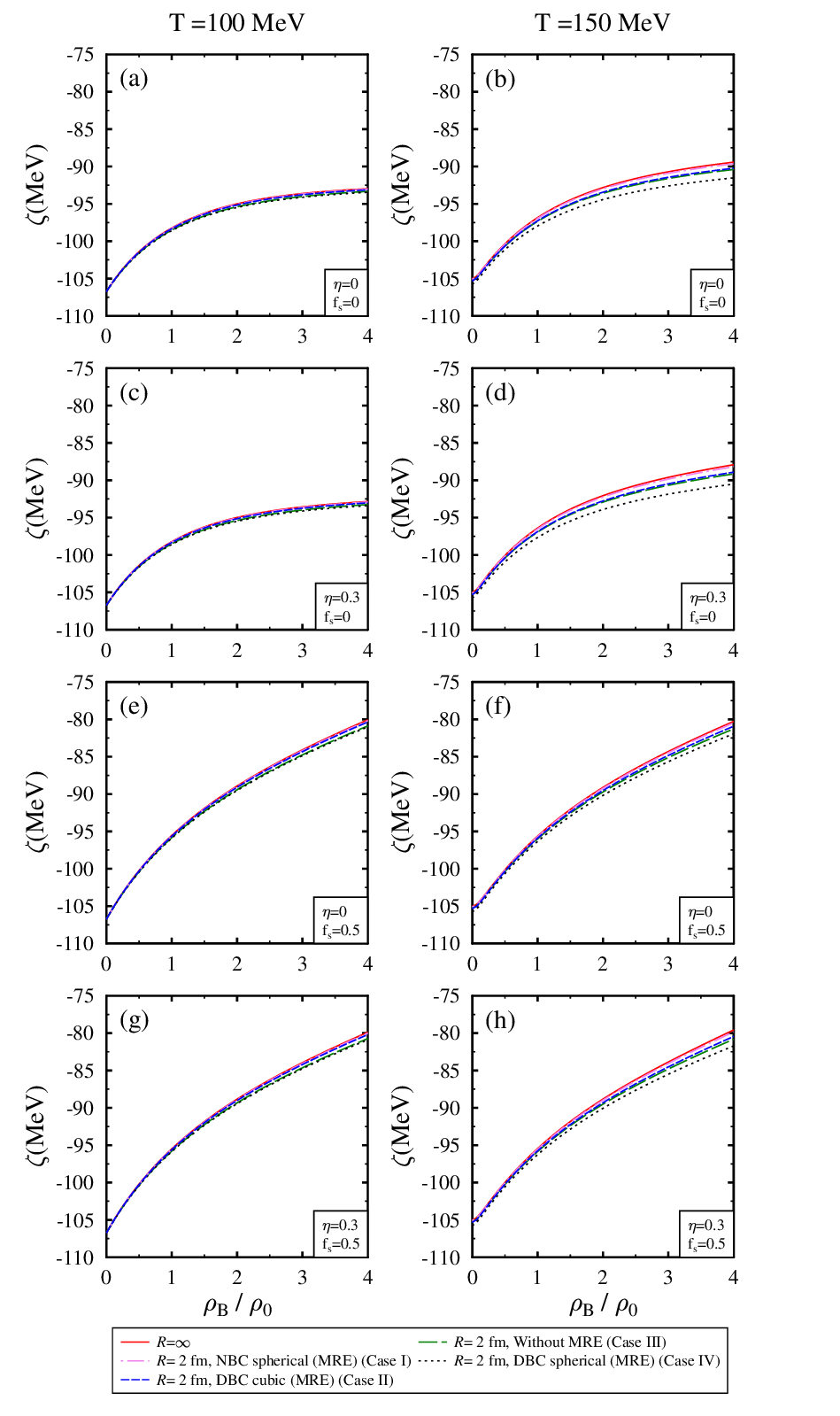}
\caption{ The variation of the scalar field $\zeta$ is shown   with respect to the baryon density $\rho_B$ (in units of nuclear saturation density $\rho_0 $), at  temperatures, $T = 100$  [in subplots (a), (c), (e) and (g)] and $150$ MeV [in subplots (b), (d), (f) and (h)].
 In the above, NBC and DBC are for Neumann and Dirichlet boundary conditions, and MRE denotes multiple reflection expansion. \label{fig:3}}
\end{figure}

The density and temperature dependence of the scalar and vector fields are obtained by solving the coupled system of equations defined through Eqs. (\ref{eq_sigma})-(\ref{eq_chi}).
 In Figs. (\ref{fig:2}) 
and (\ref{fig:3}), we plot the non-strange scalar field $\sigma$ and the strange scalar field $\zeta$ with respect to baryonic density  $\rho_B$ (in units of nuclear saturation density, $\rho_0 $) in the finite volume of hadronic medium. The results are shown for nuclear ($f_s = 0$) and strange hadronic matter
($f_s = 0.5$) at temperatures $T = 100$ and $150$ MeV and isospin asymmetry parameters, $\eta = 0$ and $0.3$. To understand the impact of finite volume, we have performed the calculations for $R = 2$ fm and compared the results at $R = \infty$, which corresponds to infinite volume hadronic matter.
For $R = 2$ fm (finite volume), we have shown the results for spherical geometry of the finite volume hadronic matter, with MRE method used to calculate the density of states under  Neumann and Dirichlet boundary conditions (shown as case (I) and (IV) in the figures).
In the figures, the results are also shown for  cubic geometry, for cubic length, $R = 2$ fm,  and considering
the density of states
calculated using
Eq. (\ref{eq_cubic_L})
(shown as case II, with Dirichlet boundary condition) and without MRE method (case III).
In case III, the finite volume of the medium is taken into account by simply applying the lower momentum cut-off, $k = \frac{\pi}{R}$. This procedure of lower momentum cut-off without using spherical and curvature effects has been used to the study of finite volume quark matter in various work \cite{Bhattacharyya:2012rp,Samanta:2017ohm}.
In the current study, we have not used the periodic and anti-periodic boundary conditions
with a discrete sum over the momentum states for the consideration of finite volume effect \cite{Abreu:2019czp}.
As has been demonstrated in Ref. \cite{Redlich:2016vvb}, the discrete sum can be transformed to a continuous integral, and the use of lower momentum cut-off in calculations of finite volume effects is justified.


For a given temperature, $T$, isospin asymmetry, $\eta$, and strangeness fraction, $f_s$, of the medium, the magnitude of scalar fields $\sigma$ and $\zeta$   decreases with an increase in the density of the medium, both for infinite ($R = \infty$) and finite ($R = 2$ fm) volume hadronic medium. In the symmetric nuclear medium, at temperature $T = 100$ MeV and $R = \infty$, the magnitude
of scalar fields $\sigma$ and $\zeta$ 
decreases by $63.6$\% and $12.9$\%,
as $\rho_B$ is increased from zero to $4\rho_0$.  
As can be seen from Figs. (\ref{fig:2}) and (\ref{fig:3}), for a given density of the medium, decreasing the value of $R$ from $\infty$ to $2$ fm leads to an increase in the magnitude of scalar fields, i.e., less drop in the magnitude of scalar fields is observed. Since the mass of baryon is directly proportional to the in-medium values of scalar fields (see Eq. (\ref{Eq_of_baryon_mass})), the effective mass of baryons will increase with a decrease in the volume of the nuclear or strange hadronic medium, at the fixed value of baryon density. From Figs. (\ref{fig:2}) 
and (\ref{fig:3}), we also observe that on moving from infinite to finite volume case, the maximum change in the scalar fields (an increase of magnitude)
is observed in the situation where finite volume calculations are performed for the spherical geometry of the medium with Dirichlet boundary conditions in the MRE method.
As the symmetric nuclear medium, at $T = 100$ MeV, is squeezed to a finite volume, with $R = 2$ fm,  values of percentage drop in the magnitude of scalar field $\sigma$, for an increase in $\rho_B$ from zero to $4\rho_0$, are observed to be  $63.2$\%, $62.5$\%, $60.9$\% and $60.2$ \%, for cases I to IV, respectively.
For the strange scalar field $\zeta$, these values of percentage drop at $R = 2$ fm are observed to be 
$12.8$\%, $12.7$\%, $12.6$\% and $12.4$ \%.
One can compare these values with the earlier quoted results for infinite nuclear matter, i.e., $R = \infty$.
In the present calculations, we are interested in the study of hadronic matter at non-zero baryon density and finite temperature. The first term of Eq.(\ref{Thermodynaics_potn}) represents the contribution of the medium with finite baryon density to the total thermodynamic potential. This is in line with the calculations of various relativistic mean field models employed to study the properties of the hadronic medium at finite baryon density \cite{Tsushima:1997df}.
The vacuum term, which is usually taken into account in the NJL/PNJL \cite{Zhao:2018nqa} and some of quark models \cite{Chahal:2023khc} and which require regularization techniques to avoid divergence, is not added in the thermodynamic potential of Eq.(\ref{Thermodynaics_potn}).
In the study of finite volume quark matter properties using the   models which 
consider the contribution of vacuum term to the thermodynamic potential,   the effective masses of quarks are observed to decrease with decrease in the volume of system \cite{MataCarrizal:2022gdr}.
On the other side, in Refs. \cite{Magdy:2019frj} quark matter properties were investigated using the quark models, neglecting  the vacuum term and an increase in the effective mass of quarks is observed for a decrease in the volume . 

\begin{figure}[htbp]
\centering
\includegraphics[width=0.75\textwidth]{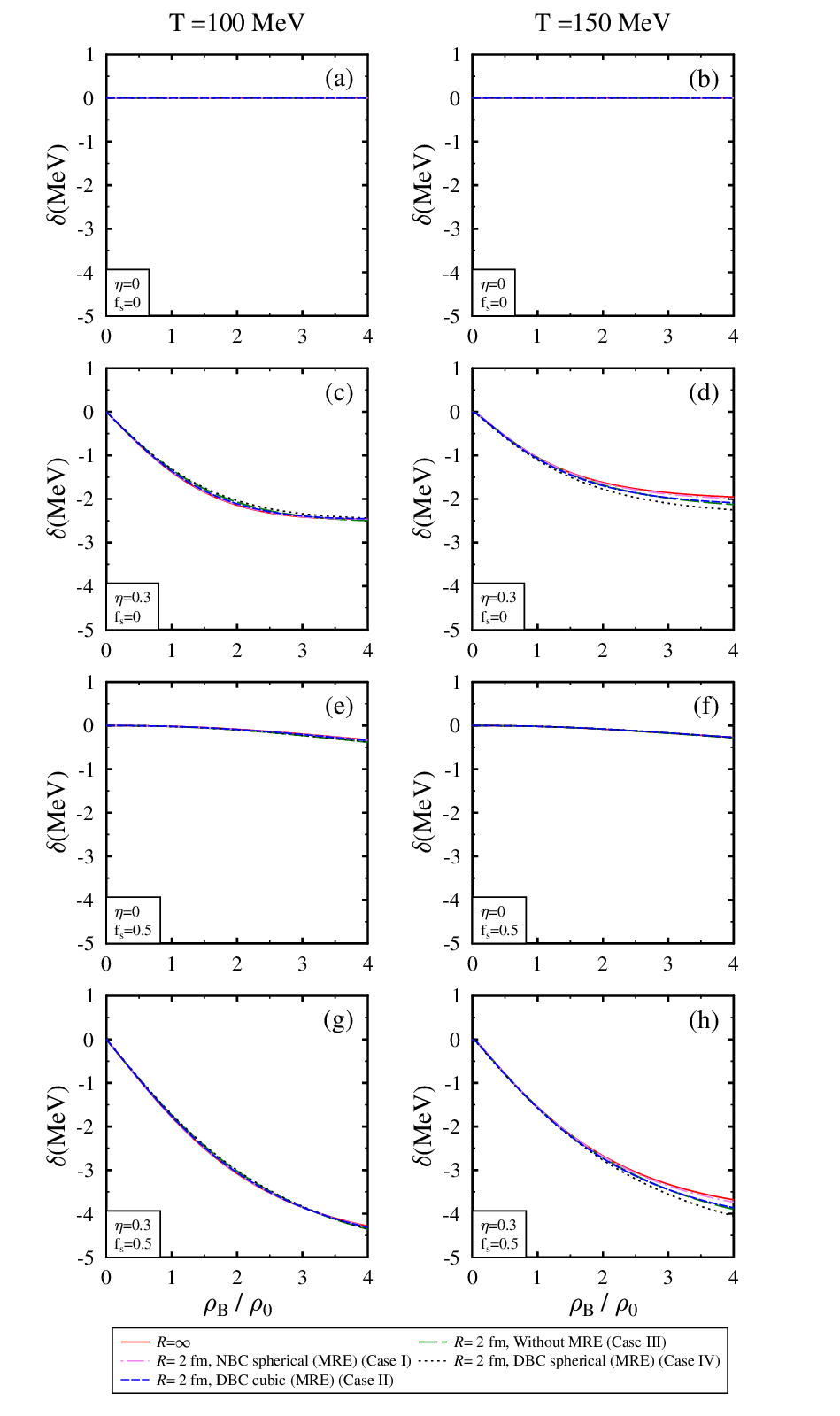}
\caption{ The variation of scalar field $\delta$ is shown   with respect to baryon density $\rho_B$ (in units of nuclear saturation density $\rho_0 $), at  temperatures, $T = 100$  [in subplots (a), (c), (e) and (g)] and $150$ MeV [in subplots (b), (d), (f) and (h)].
%
In the above, NBC and DBC are for Neumann and Dirichlet boundary conditions, and MRE denotes multiple reflection expansion.  \label{fig:4}}
\end{figure}

\begin{figure}[htbp]
\centering
\includegraphics[width=0.75\textwidth]{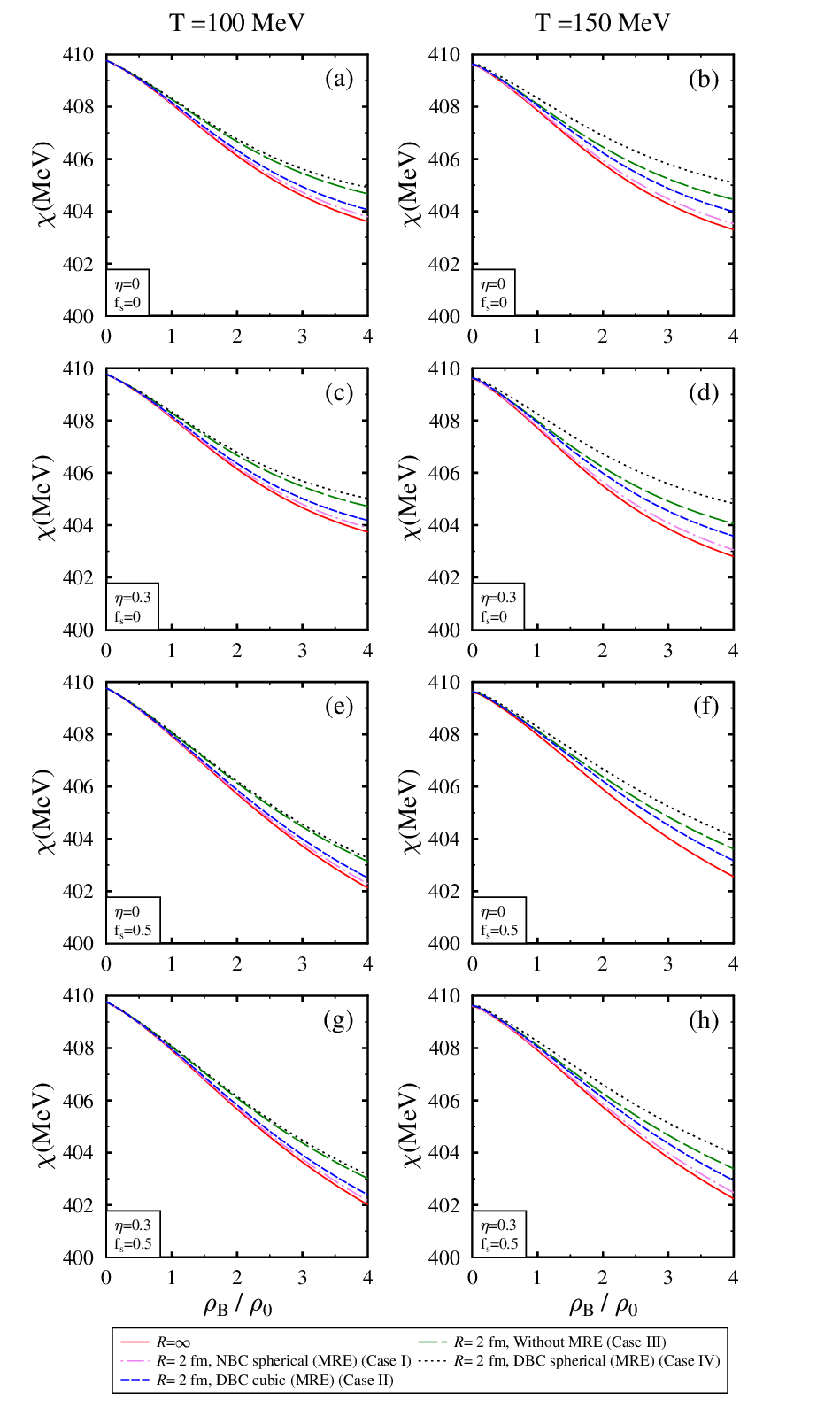}
\caption{ The variation of scalar dilaton field $\chi$ is shown   with respect to baryon density $\rho_B$ (in units of nuclear saturation density $\rho_0 $), at  temperatures, $T = 100$  [in subplots (a), (c), (e) and (g)] and $150$ MeV [in subplots (b), (d), (f) and (h)].
%
 In the above, NBC and DBC are for Neumann and Dirichlet boundary conditions, and MRE denotes multiple reflection expansion.  \label{fig:5}}
\end{figure}

An increase in the isospin asymmetry of the medium from zero to finite value causes less decrease in the magnitude of scalar fields $\sigma$ and $\zeta$.
The scalar field $\sigma$ which is proportional to light quark condensates is observed to affect more compared to the scalar field $\zeta$, which is proportional to the strange quark condensate. The impact of isospin asymmetry is observed to be more in the infinite volume matter compared to the finite volume case. For example, 
increasing the value of isospin asymmetry parameter $\eta$ from zero to 0.3,  
magnitude of scalar field $\sigma$ increases by $1.63$\%  and $1.44$\% at $R = \infty$ and $2$ fm (case IV), respectively, at baryon density $\rho_B = 4\rho_0$ and
temperature, $T = 100$ MeV.
The finite strangeness fraction of the medium impacts more the strange scalar field $\zeta$ compared to the non-strange scalar field $\sigma$. 
For a given density and temperature of the medium, increasing the strangeness fraction, $f_s$, from $0$ to $0.5$ causes an increase in the
magnitude of the scalar field $\sigma$ and a decrease in $\zeta$.  In isospin symmetric medium, for $T = 100$ MeV and  $\rho_B = 4\rho_0$, as $f_s$ is increased from $0$ to $0.5$, at $R = \infty$ (2 fm  (case IV)), the magnitude of scalar field $\sigma$ increases by $2.8$\% ($1.1$\%), whereas, the magnitude of $\zeta$ field decreases by $1.38$\% ($1.33$\%). As can be seen, the impact of strangeness is observed to be higher in the infinite strange matter.
In Figs. (\ref{fig:4}) and (\ref{fig:5}), the variation of the scalar isovector field $\delta$ and the scalar dilaton field $\chi$  is shown, respectively. The scalar isovector field $\delta$ contributes most significantly in the asymmetric nuclear and hyperonic medium. The dilaton field $\chi$ is observed to have small variations as a function of density of the medium which is the reason that this field is treated as a frozen glue ball in some studies \cite{Papazoglou:1998vr}.
Decreasing the value of system size from $R = \infty$ to $2$ fm enhances the value of the dilaton field, $\chi$, in the medium, i.e., smaller drop is observed in finite volume in comparison to the infinite volume case.

\subsection{Kaon and antikaon in finite volume matter} \label{method6}

\begin{figure}[htbp]
\includegraphics[width=0.75\textwidth]{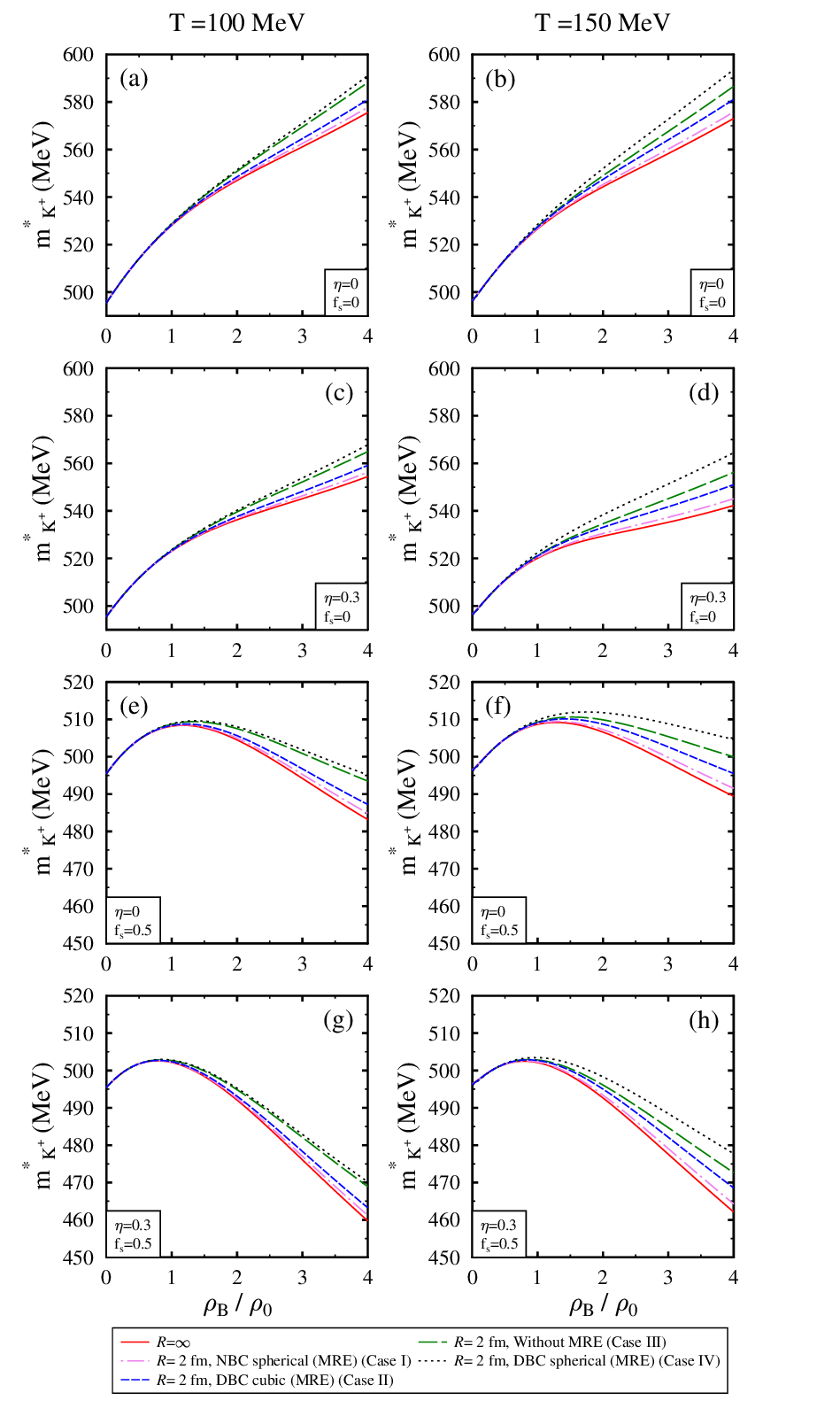}
\caption{The in-medium masses of $K^+$ mesons are plotted with respect to the baryon density $\rho_B$ (in units of nuclear saturation density $\rho_0 $), at
 temperatures, $T = 100$  [in subplots (a), (c), (e) and (g)] and $150$ MeV [in subplots (b), (d), (f) and (h)].
In the above, NBC and DBC are for Neumann and Dirichlet boundary conditions, and MRE denotes multiple reflection expansion. \label{fig:6}}
\end{figure}

\begin{figure}[htbp]
\includegraphics[width=0.75\textwidth]{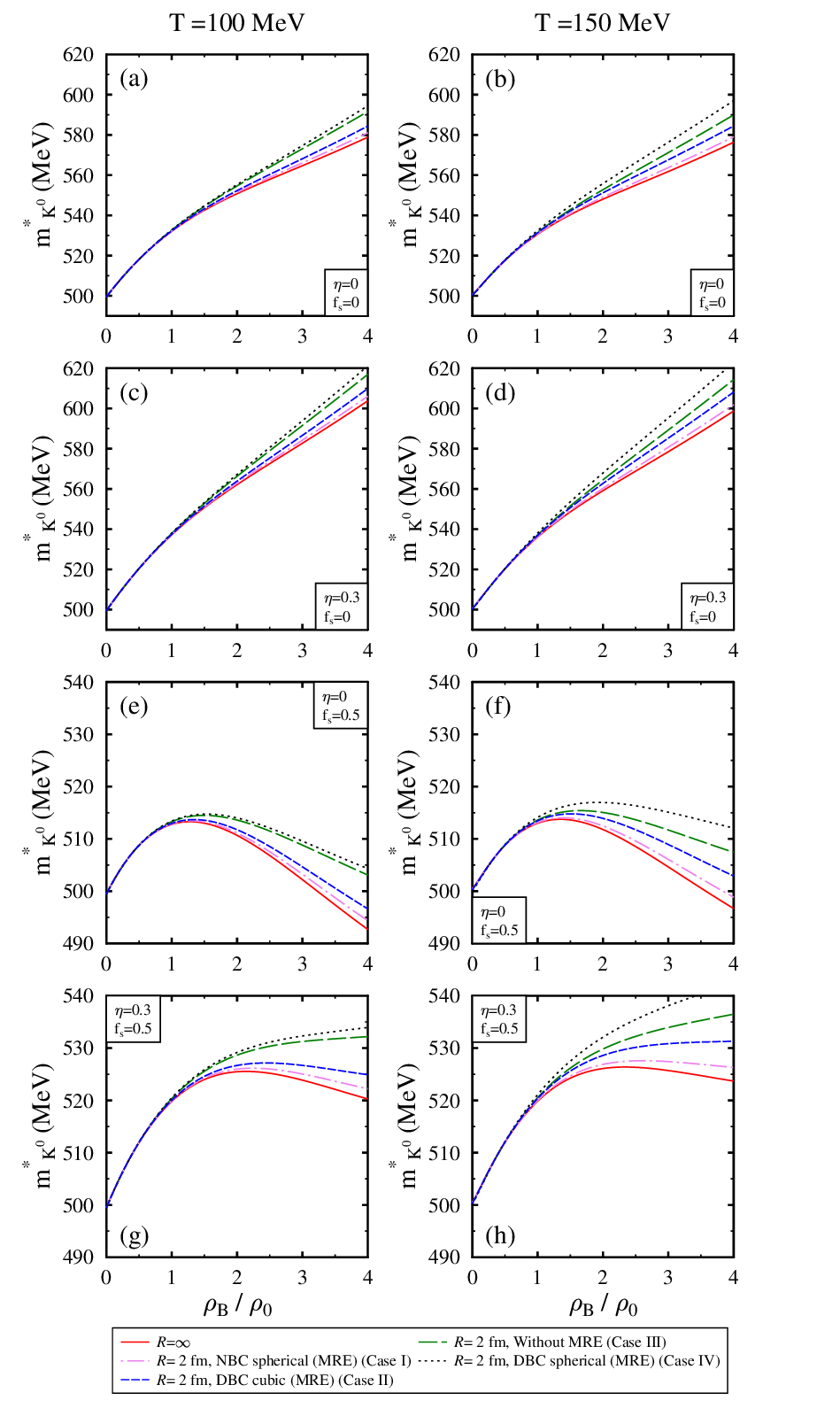}
\caption{The in-medium masses of $K^0$ mesons are plotted  with respect to the baryon density $\rho_B$ (in units of nuclear saturation density $\rho_0 $), at
 temperatures, $T = 100$  [in subplots (a), (c), (e) and (g)] and $150$ MeV [in subplots (b), (d), (f) and (h)].
%
 In the above, NBC and DBC are for Neumann and Dirichlet boundary conditions, and MRE denotes multiple reflection expansion. \label{fig:7}}
\end{figure}

\begin{figure}[htbp]
\includegraphics[width=0.75\textwidth]{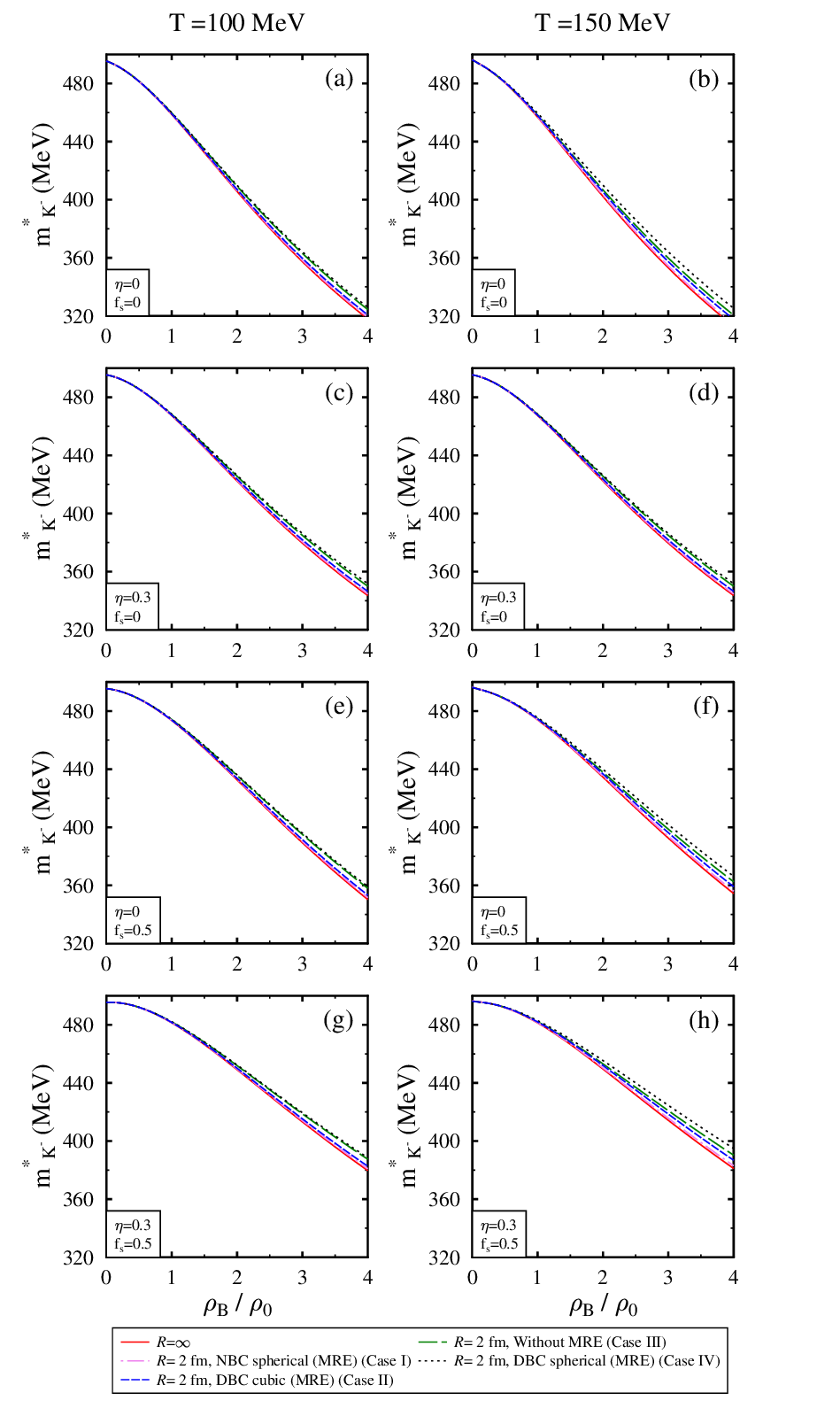}
\caption{The in-medium masses of $K^-$ mesons are plotted as a function of the baryon density $\rho_B$ (in units of nuclear saturation density $\rho_0 $), at
 temperatures, $T = 100$  [in subplots (a), (c), (e) and (g)] and $150$ MeV [in subplots (b), (d), (f) and (h)].
%
 In the above, NBC and DBC are for Neumann and Dirichlet boundary conditions, and MRE denotes multiple reflection expansion. \label{fig:8}}
\end{figure}

\begin{figure}[htbp]
\includegraphics[width=0.75\textwidth]{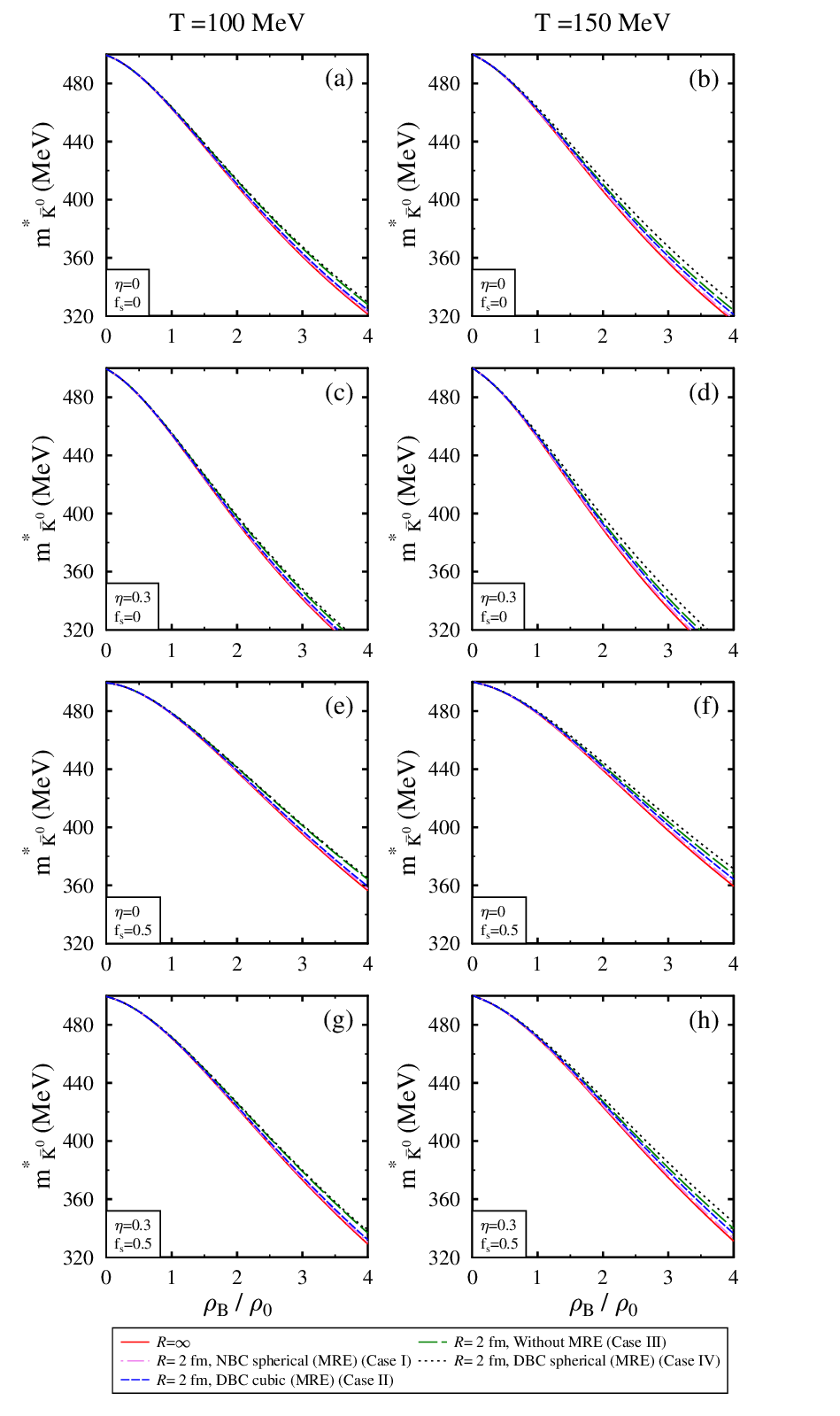}
\caption{The in-medium masses of $\bar{K^0}$ mesons are plotted as a function of the baryon density $\rho_B$ (in units of nuclear saturation density $\rho_0 $)
, at
 temperatures, $T = 100$  [in subplots (a), (c), (e) and (g)] and $150$ MeV [in subplots (b), (d), (f) and (h)].
%
 In the above, NBC and DBC are for Neumann and Dirichlet boundary conditions, and MRE denotes multiple reflection expansion.  \label{fig:9}}
\end{figure}

\begin{figure}[htbp]
	\centering
	\includegraphics[width=0.73\textwidth]{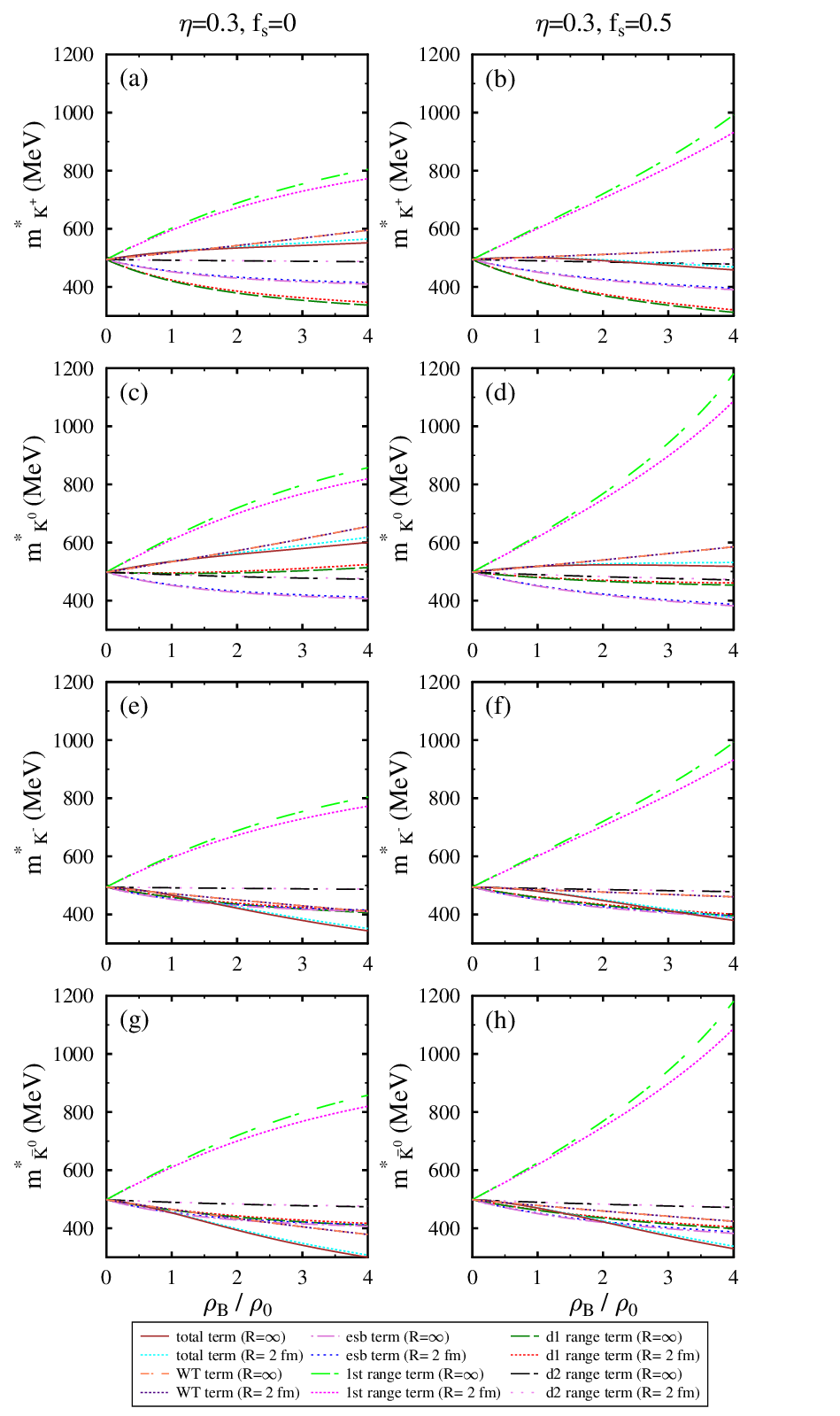}
	\caption{The contributions  of the individual terms of interaction Lagrangian to the in-medium masses of $K$ and $\bar{K}$ mesons are plotted as a function of   baryon density $\rho_B$ (in units of nuclear saturation density $\rho_0 $) in asymmetry ($\eta = 0.3$) nuclear ($f_s = 0$) 
	[in subplots (a), (c), (e) and (g)]  and strange 
	($f_s = 0.5$)
	[in subplots (b), (d), (f) and (h)] matter, at $T = 100$ MeV.
	In figure, esb and WT denote explicit symmetry breaking and Weinberg Tomozawa, respectively.   \label{fig:term_com}}
\end{figure}

In this section, we discuss the medium modification of kaon ($K^+, K^0$) and antikaon ($K^-, \bar{K^0}$)  masses in the strange hadronic matter with finite volume. The in-medium masses of $K$ and $\Bar{K}$ mesons are calculated using the medium modified values of vector and scalar densities of baryons as well as
the scalar fields $\sigma,\zeta$ and $\delta$, in the expressions of self energies, as given in Eqs. (\ref{Eq_dis_kaon}) and  (\ref{Eq_dis_antikaon}). Figures (\ref{fig:6}) and (\ref{fig:7}) show the medium modified masses of $K^+$ and $K^0$  mesons with respect to the baryonic density $\rho_B$ (in units of nuclear saturation density, $\rho_0$) at temperatures $T = 100$ and 150 MeV, for the system sizes $R = \infty$ and $2$ fm.
 In the nuclear medium, both for $R = \infty$  and $2$ fm, the in-medium masses of kaons, $K^+$ and $K^0$, increase as a function of baryon density, $\rho_B$, of the nuclear medium, i.e., both members of $K$-isospin doublet feel repulsive interactions.
 On the otherside, in the case of antikaon doublet $\bar{K}$, the in-medium masses of $K^{-}$ and $\bar{K^{0}}$, plotted in 
  Figs. (\ref{fig:8}) and (\ref{fig:9}), decrease as a function of density
  of the medium.
 As can be seen from the interaction Lagrangian density given by Eq.(\ref{Eq_interaction_KB}), the medium modification of kaons and antikaons within the chiral SU(3) hadronic mean field model occurs through the contribution of Weinberg Tomozawa term, the explicit symmetry breaking term and three range terms (1st  range term stemming from kinetic term of pseudoscalar mesons and the other two from $d_1$ and $d_2$ terms, respectively).
 In  Fig. (\ref{fig:term_com}), contributions of individual terms to the effective masses of kaons and antikaons are shown for finite volume matter, $R = 2$ fm (Dirichlet boundary conditions with MRE case) and are compared with the results calculated in the infinite volume matter, i.e., $R = \infty$.
 The results are shown for temperature, $T = 100$ MeV and isospin asymmetry, $\eta = 0.3$.
 The left panel shows the results for the nuclear medium ($f_s = 0)$,
 whereas on the right side, strangeness fraction $f_s$ is fixed at 0.5.
  In the nuclear medium, as a function of the baryon density $\rho_B$, the Weinberg Tomozawa term gives repulsive contributions to the in-medium masses of kaons and attractive to antikaons,  leading to an increase and a decrease in their in-medium masses, respectively. The explicit symmetry-breaking term of the model provides attractive contributions to both, kaons and antikaons, thereby reducing their in-medium masses as a function of $\rho_B$, for both $R = \infty$ and $2$ fm. The first range term gives repulsive contributions to both, kaons and antikaons, whereas the $d_1$ and $d_2$ terms contribute to the attractive interactions for the in-medium modification in the masses of kaons and antikaons,
 (the term $d_2$ is less attractive compared to the $d_1$ terms, for the $K^+$ and $K^-$ mesons, as we can notice in Fig. (\ref{fig:term_com})). As a combined effect, the repulsive interaction for $K^{+}$ and $K^{0}$ mesons, mainly from the WT term and the first range term, dominate over the attractive interactions provided by the explicit symmetry breaking term, the \(d_1\) and \(d_2\) terms. This results in a net increase in their in-medium masses as a function of the density of the nuclear medium. The opposite is true for the antikaon doublet, i.e., the attractive interactions provided by the WT term, explicit symmetry breaking term, and the \(d_1\) and \(d_2\) terms
  for antikaons dominate over the repulsive interactions from the first range term. This causes a net decrease in their in-medium masses with increasing density of the nuclear matter.

For a given density of the nuclear medium, when the value of $R$ is decreased from $\infty$ to $2$ fm, 
the in-medium mass of kaons and antikaons increases. For example, in symmetric nuclear matter, for baryon density $\rho_B$ = $4\rho_0$ and temperature $T = 100$ MeV, at $R = \infty$ (2 fm, Dirichlet boundary conditions with spherical geometry and MRE method),
the effective masses of $K^{+}$,  
 $K^{0}$,  $K^{-}$ and $\bar{K^{0}}$
 are observed to be $575.56$ $(591.06)$, $578.88$ ($594.47$), $317.96$ $(326.00)$ and $321.28$ $(329.40)$ MeV, respectively.
 Looking at the behavior of individual terms as a function of $R$,  (see Fig. (\ref{fig:term_com})), effective mass increases with a decrease in $R$ for all terms except for the first range term, for which  the
 effective mass decrease with decreasing $R$.
 For a given density of the nuclear matter, an increase of the isospin asymmetry parameter from zero to finite value causes the mass splitting between the members of a given isospin doublet of kaons and antikaons. At baryon density $\rho_B = 4\rho_0$,  as $\eta$ is changed from zero to $ 0.3$, in infinite nuclear matter ($R = \infty$), the mass of $K^{+}$ meson decreases by $21.16$ MeV, whereas the effective mass of $K^{0}$ increases by $24.89$ MeV.
 In case of antikaon doublet $\bar{K}$, the effective mass of $K^{-}$ ($\bar{K^0}$) increases (decreases) by $25.6$ ($21.85$) MeV, as $\eta$ is changed from $0$ to 0.3.
 When the medium is confined to finite volume  with $R = 2$ fm (case IV), the values of mass-shift, as a function of isospin asymmetry
 (changing $\eta$ from $0$ to $0.3$), for $K^{+}$, $K^{0}$, $K^{-}$ and $\bar{K^0}$
mesons are observed to be $-23.28$, $26.66$, $25.73$ and $-22.06$ MeV, respectively.
From the above, we observe that the magnitude of  mass splitting between the members of a given isospin doublet increases with decease in the system size, i.e., changing the volume of medium from infinite to finite value.

 Now we discuss the impact of strangeness fraction, $f_s$, on the effective masses of kaons and antikaons in the infinite and finite strange hadronic medium.
As can be seen from Figs. (\ref{fig:6}) and   (\ref{fig:7}),  the effective masses of  $K^+$ and $K^0$ mesons first increase upto certain density beyond which  a decrease  is observed with further increase in the baryon density, $\rho_B$, of the strange medium, except for the mass of $K^0$ mesons at $\eta = 0.3, f_s = 0.5$ in the finite volume case.
 As discussed earlier, the finite isospin asymmetry of the medium causes mass splitting between the isospin partners $K^{+}$ and $K^{0}$ mesons, and the effective mass of $K^{+}$ decreases whereas the mass of $K^{0}$ increases with $\eta$. Thus, at finite strangeness fraction $f_s$ and isospin asymmetry $\eta = 0.3$, the baryon density upto which the mass of $K^+$ mesons  increases, decrease to a lower value compared to the symmetric matter ($\eta = 0$) case. On the otherside, increase of isospin asymmetry decrease the impact of  strangeness fraction on the masses of $K^{0}$ mesons, i.e., less decrease in the mass of $K^{0}$ mesons is observed due to strangeness fraction $f_s$, with increase of $\eta$ from zero to finite value [as can be seen on comparing the subplots (g) and (h) with subplots (e) and (f) of Fig. (\ref{fig:7})].
 As can be seen from Fig. (\ref{fig:term_com}), on transition from non-strange to a strange medium,
 for the case of kaons, the strength of repulsive contributions of Weinberg term decreases and the attractive contributions of explicit symmetry breaking term and the  $d_1$ term
 increase, resulting in overall decrease in the effective mass of $K^{+}$ and $K^{0}$ mesons at higher values of baryon density of strange medium.
  Consideration of the finite volume of the strange medium decreases the impact of attractive interactions, i.e., in the strange medium also, the effective mass of kaons and antikaons increases on changing $R$ from $\infty$ to $2$ fm. 
  An increase of strangeness fraction causes a lesser drop in the masses of $K^-$ and $\bar{K^0}$ mesons, i.e., the effective masses
  of antikaons increase with the strangeness fraction of the medium (see Figs. (\ref{fig:8}) and   (\ref{fig:9})). 
  To quote in terms of numbers, as the 
  strangeness fraction $f_s$ changes from $0$ to $0.5$, at baryon density $4\rho_0$ and temperature $T = 100$ MeV, the mass shift for
  $K^{+}$, $K^{0}$, $K^{-}$ and $\bar{K^0}$ are observed to be $-92.39$ ($-95.93$), $-95.71$ ($-90.15$), $32.35$ ($33.21$) and $35.15$ ($35.75$) MeV, respectively, at $R = \infty$ ($2$ fm, (case IV)). As can be 
 seen from Figs. (\ref{fig:6}) to (\ref{fig:9}),
  among the different cases at finite volume, the effective masses of kaons and antikaons modify most significantly in the case when the density of states is calculated using the MRE method and Dirichlet boundary conditions are implemented, whereas the effects of Neumann boundary conditions cause the least change.
 In Ref. \cite{Zhao:2018nqa}, the effect of finite volume on the masses of pseudoscalar mesons was studied using the PNJL model
 and the masses of kaons are observed to increase with a decrease in the volume of the system.

\subsection{Medium modified mass and decay width of $\Phi$ meson} \label{method7}
\begin{figure}[h]
	\includegraphics[width=0.72\textwidth]{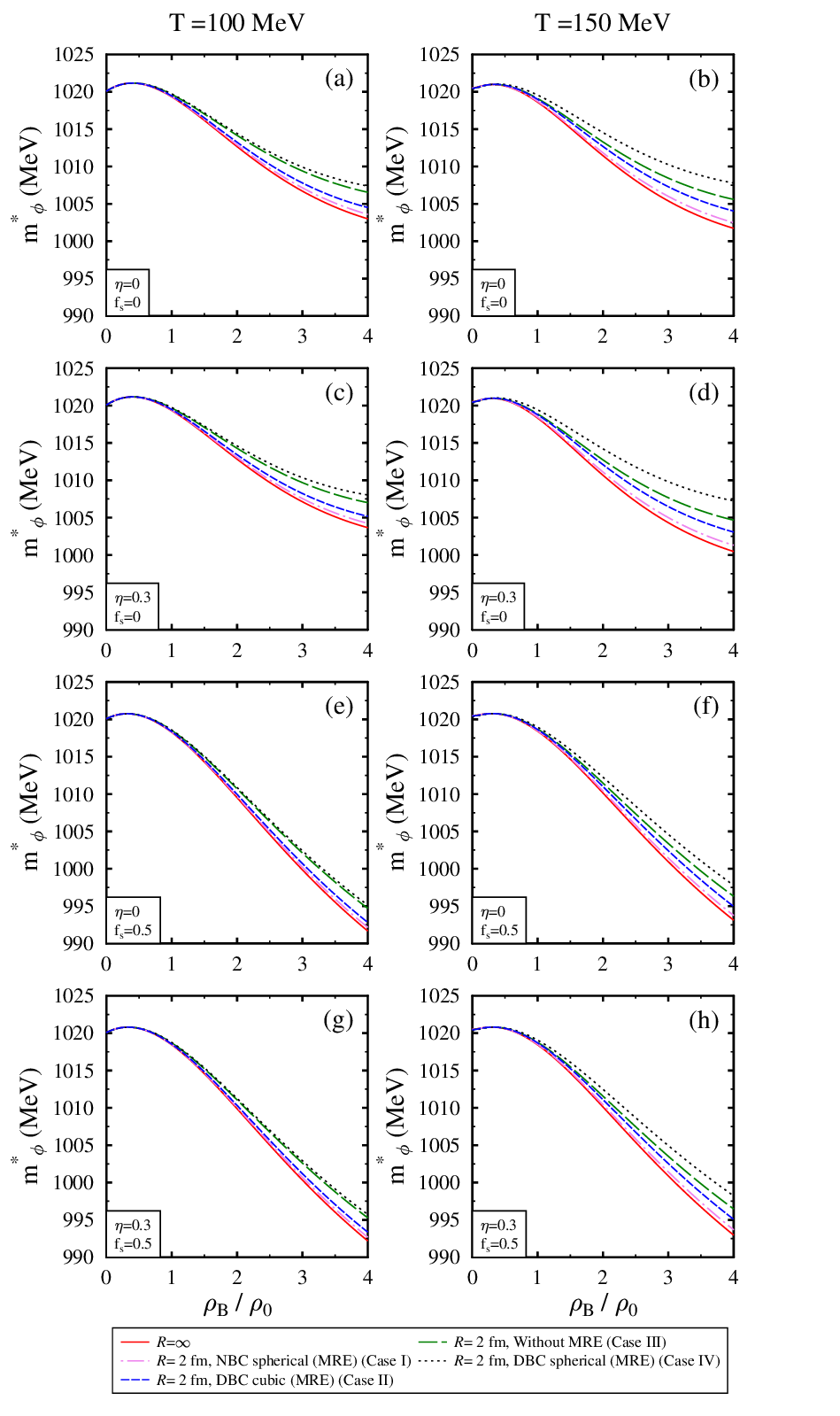}
	\caption{The in-medium masses of $\phi$ mesons
	are plotted as a function of the baryon density $\rho_B$ (in units of nuclear saturation density $\rho_0 $)
	, at
	 temperatures, $T = 100$  [in subplots (a), (c), (e) and (g)] and $150$ MeV [in subplots (b), (d), (f) and (h)].  
 In the above, NBC and DBC are for Neumann and Dirichlet boundary conditions, and MRE denotes multiple reflection expansion. \label{fig:10}}
\end{figure}

In this subsection, we shall discuss the
impact of medium modified masses of kaons and antikaons on the masses and decay widths of $\phi$ mesons in finite volume nuclear and strange hadronic matter. 
As discussed in Sec. \ref{method4}, the self-energy of $\phi$ mesons is calculated at one loop level for the process $\phi \rightarrow  K \bar{K}$.
The effective mass of $\phi$ mesons in the finite volume strange hadronic matter is calculated using Eq. (\ref{33}), which depends upon the real part of the self-energy of $\phi$ mesons, given by Eq. (\ref{real_part_of_self_eng_phi}). The medium modified masses of kaons and antikaons calculated in the finite volume strange hadronic matter are plugged
into this equation to obtain the effective mass of $\phi$ mesons.  In Fig. (\ref{fig:10}), we show the variation of the effective mass, $m_{\phi}^{*}$, of the $\phi$ mesons in the nuclear and strange hadronic medium for the infinite ($R = \infty$) and finite ($R = 2$ fm) volume of hadronic matter, considering the
cut-off parameter $\Lambda_c$= 3 GeV, to regularize the loop integral of the real part of self-energy.
We observe that the effective mass of $\phi$ mesons first increase slightly and then decreases with the increase in the baryonic density of the medium, for both $R =\infty$ and $2$ fm. For a given density, a decrease in the volume of the medium causes an increase in the mass of these mesons. For example, in the nuclear matter at baryon density $\rho_B = 4\rho_0$ and temperature $T = 100$ MeV, the effective masses of $\phi$ mesons are observed to be $1003.0$ ($1003.7$) and $1007.4$ ($1008.0$) MeV, at $R = \infty$ and $2$ fm, case IV, for isospin asymmetry $\eta = 0 (0.3)$.
In the strange hadronic medium, with $f_s = 0.5$, these in-medium masses of $\phi$ mesons change to  
$991.67$ ($992.19$) and $995.13$ ($995.73$) MeV, at $R = \infty$ and $2$ fm , for  asymmetry parameter $\eta = 0 (0.3)$.
As the volume of the medium is changed from infinite to a finite value, the percentage increase in the effective mass of $\phi$ mesons is observed to be more in the nuclear medium compared to the strange hadronic medium. 
Also, as was the case previously, the impact of finite volume is observed to be more with the spherical geometry and under Dirichlet boundary conditions. In Fig.(\ref{fig:10}), results are shown for the fixed value of the cut-off parameter, $\Lambda_{c}$. For given density and temperature of the medium, an increase in the value of
cut-off parameter $\Lambda_c$ causes an increase in the effective mass of these mesons \cite{Kumar:2020vys}.
Also, an increase in the temperature of the nuclear and strange hadronic medium will cause less drop in the mass of $\phi$ mesons. 


In  Refs. \cite{Cobos-Martinez:2017woo,Cobos-Martinez:2017vtr}, 
masses and decay widths of $\phi$ mesons were calculated in the symmetric infinite nuclear matter through the mass modification of kaons and antikaons appearing in the loop integral. In these studies, the effective mass of kaons is calculated using the QMC model, and both kaons and antikaons were considered to have equal mass in the nuclear medium.
However, as discussed in the previous section, within the chiral SU(3) model, kaons and antikaons are modified differently in the hadronic medium. This 
causes significantly different mass modifications of $\phi$ mesons in the present calculations within infinite and finite nuclear matter compared to the Ref.
\cite{Cobos-Martinez:2017woo,Cobos-Martinez:2017vtr}.
A very modest decline in the in-medium mass of the $\phi$ meson in the asymmetric strange matter at zero temperature was noticed in Ref. \cite{Mishra:2014rha} through the use of QCD sum rules \cite{Drukarev:1988kd} and the unification of the chiral SU(3) model.

\begin{figure}[htbp]
	\includegraphics[width=0.72\textwidth]{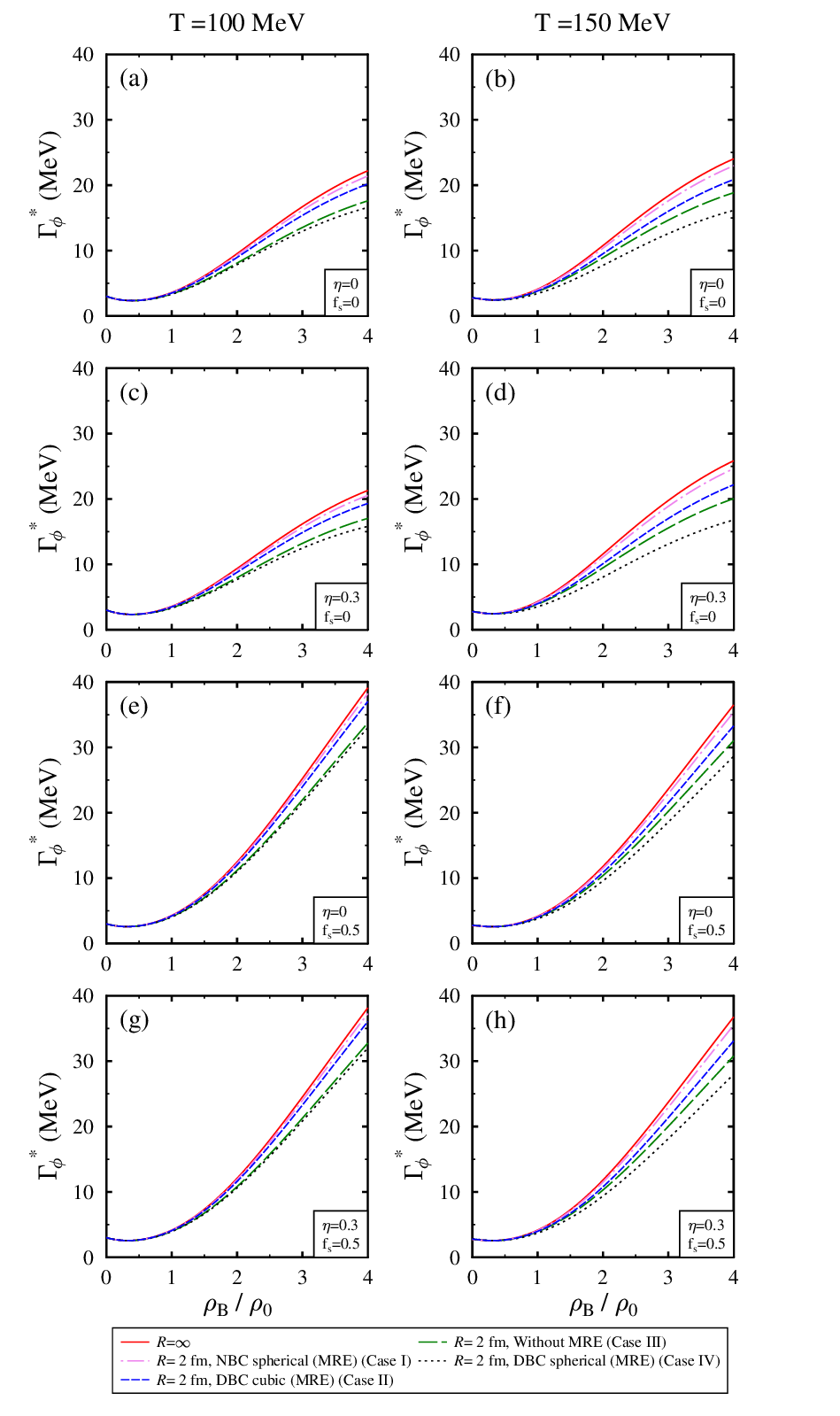}
	\caption{The in-medium decay widths of $\phi$ mesons 
 are plotted as a function of the baryon density $\rho_B$ (in units of nuclear saturation density $\rho_0 $)
 , at
  temperatures, $T = 100$  [in subplots (a), (c), (e) and (g)] and $150$ MeV [in subplots (b), (d), (f) and (h)].
	    In the above, NBC and DBC are for Neumann and Dirichlet boundary conditions, and MRE denotes multiple reflection expansion. \label{fig:11}}
\end{figure}

In Fig. (\ref{fig:11}), the medium modified decay widths, $\Gamma_{\phi}^{\star}$, of the $\phi$ mesons, decaying to $K$$\Bar{K}$ pairs and calculated using 
Eq. (\ref{phi_decay_width}), are plotted as a function of baryon density $\rho_B$, for $\Lambda_c$ = 3 GeV. 
As can be seen from the figure, the $\phi$ meson decay width increases with increase in the baryonic density, $\rho_B$, in both infinite and finite hadronic medium.
However, for a given density of the medium, as the value of system size, $R$, decreases to a finite value, due to the increase of effective mass of $K$ and $\bar{K}$ mesons, the decay width decreases, i.e., less phase space is available for the decay process $\phi \rightarrow K \bar{K}$,  as the volume of the hadronic medium becomes finite. At baryon density $4 \rho_0$, the decay widths of $\phi$ mesons are observed to be $22.20$ ($21.29$) and $16.58$ ($15.80$)  MeV at $R = \infty$ and $2$ fm, case IV, for $\eta = 0 (0.3)$.
Moving to the strange matter ($f_s = 0.5$), 
decay widths of $\phi$ mesons for
$\eta = 0 (0.3)$ are observed to be
$39.09$ ($38.15$) and $32.98$ ($31.99$)  MeV at $R = \infty$ and $2$ fm, respectively.
Although the masses and decay width of $\phi$ mesons have been studied in the infinite nuclear matter \cite{Hatsuda:1991ez,Hatsuda:1996xt,Oset:2000eg,Gubler:2015yna,Klingl:1997tm}, the current study is the first attempt to study the impact of finite volume. 


\section{Summary} \label{summary}

To summarize, in the present work, we investigated the
mass modifications of kaons and antikaons in the isospin asymmetric strange hadronic matter confined to the
finite volume using the chiral SU(3) hadronic mean field model.
The finite volume effects are introduced in the present work
through the calculations of the density of states using the MRE method and applying the lower momentum cut-off in the expressions of scalar and vector densities of baryons in terms of system size $R$. 
The magnitude of the scalar fields $\sigma$ and  $\zeta$, which modify the properties of baryons within the chiral SU(3) model, are observed to increase as the volume of the medium is considered as finite,i.e., $R$ is changed from infinite to a finite value.
This increases the effective mass of kaons and antikaons with a decrease of $R$. While calculating the in-medium modifications of kaons, antikaons, and $\phi$ meson properties in the finite volume of medium, calculations are done considering (i) Neumann boundary conditions in the MRE method and spherical geometry, (ii)  Dirichlet boundary condition with MRE method and considering spherical and cubic geometry and (iii) applying only the lower momentum cut-off, $\pi/R$, in the expressions of scalar and vector densities of baryons without calculating the density of states using MRE method. 
Among these different cases, the most significant modifications
occur when calculations are done considering the spherical geometry of the finite volume of the medium and applying Dirichlet boundary conditions.

Employing the in-medium masses of $K$ and $\Bar{K}$ mesons calculated using the chiral SU(3) model,  the masses and decay width of the $\phi$ mesons are also calculated in the finite volume nuclear and hyperonic medium. The effective
masses of $\phi$ mesons are observed to increase, whereas the decay width decreases as the value of $R$ is changed from infinite to finite value.
%
  The impact of finite volume, using periodic-antiperiodic boundary conditions, on the in-medium properties of hadrons will be investigated in the future work. Further, the impact of coupled channel calculations, considering finite volume, on the in-medium masses of different mesons will also be investigated in the future.
Present studies of $\phi$ mesons will also be of relevance in understanding the observables from the experiments, for example, 
J-PARC E16, dedicated to investigate the $\phi$
meson mass shift from the measurement of di-electron spectra in $p-A$ collisions \cite{Jparc}.
Additionally, a proposal has been made at J-Lab to investigate the interaction of helium nuclei with $\phi$  mesons after the 12 GeV upgrade \cite{E16}
and in-medium mass-shift and decay width will be useful to explore the possibility of $\phi$ mesic nuclei in such experiments.


\end{document}